\documentclass[aps,prb,preprint,amssymb]{revtex4}

\usepackage{dcolumn}
\usepackage{bm}
\usepackage{graphicx,color}
\usepackage{epstopdf}
\usepackage{xspace}
\usepackage{amsmath}
\usepackage[english]{babel}
\usepackage{verbatim}
\usepackage{braket}
\usepackage{booktabs}
\usepackage[version=3]{mhchem}

\newcommand{\ac}{$\alpha_\mathrm{c}$}
\newcommand{\lc}{$\lambda_\mathrm{c}$}
\newcommand{\ain}{$\alpha_\mathrm{a^*}$}

\newcommand{\ali}{$\alpha_{\mathrm{i}}$}
\newcommand{\aimag}{$\alpha_\mathrm{i,mag}$}

\newcommand{\TN}{$T_\mathrm{N}$}
\newcommand{\TNOne}{$T_\mathrm{N1}$}
\newcommand{\TNTwo}{$T_\mathrm{N2}$}
\newcommand{\GPS}{Gd$_2$PdSi$_3$}
\newcommand{\LPS}{Lu$_2$PdSi$_3$}
\newcommand{\CPS}{Ce$_2$PdSi$_3$}

\newcommand{\cp}{$c_{\mathrm{p}}$}
\newcommand{\cpt}{$c_{\mathrm{p}}$/T}
\newcommand{\cptmag}{$c_{\mathrm{p,mag}}$/T}
\newcommand{\cpmag}{$c_{\mathrm{p,mag}}$}

\newcommand{\dchicA}{$\Delta \chi_{\mathrm{c}}^{\mathrm{A}}$}

\newcommand{\jmk}{J/(mol K)}

\newcommand{\beginsupplement}{%
        \setcounter{table}{0}
        \renewcommand{\thetable}{S\arabic{table}}%
        \setcounter{figure}{0}
        \renewcommand{\thefigure}{S\arabic{figure}}%
     }

\begin{document}
	
	\title{Magnetoelastic Coupling and Phases in the Skyrmion Lattice Magnet \ce{Gd2PdSi3} Discovered by High-resolution Dilatometry}
	\author{S. Spachmann$^{1}$}
	\email{sven.spachmann@kip.uni-heidelberg.de}
	\author{A. Elghandour$^{1}$}
	\author{M. Frontzek$^{2,4}$}
	\author{W. L\"{o}ser$^{3}$}
	\author{R. Klingeler$^{1,4}$}
	
	\affiliation{$^1$Kirchhoff Institute for Physics, Heidelberg University,Germany}
	\affiliation{$^2$Oak Ridge National Laboratory, Oak Ridge, USA}
	\affiliation{$^3$Leibniz Institute for Solid State and Materials Research (IFW), Dresden, Germany}
	\affiliation{$^4$Institute of Solid State and Materials Physics, Dresden University of Technology, Germany}
	\affiliation{$^4$Centre for Advanced Materials (CAM), Heidelberg University, Germany}
	\date{\today}

\begin{abstract}
    We report detailed thermodynamic studies on high-quality single crystals of the centrosymmetric skyrmion-hosting intermetallic \GPS\ by means of high-resolution capacitance dilatometry in fields up to 15~T which are complemented by specific heat and magnetization studies.
    Our dilatometric measurements show magnetoelastic effects associated with antiferromagnetic order at \TNOne~=~22.3~K and \TNTwo~=~19.7~K, as well as strong field effects in an applied magnetic field of 15~T up to 200~K (150~K) for $B\parallel c$ ($B\parallel a$*, i.e. $B\perp c$). The data allow us to complete the magnetic phase diagram, including a new feature at $T^*\approx 12$~K below which a new degree of freedom becomes relevant. For the first time, the magnetic B vs. T phase diagram for the $a$*-axis is also reported. Gr\"{u}neisen analysis shows the onset of magnetic contributions around 60~K, i.e., well above \TNOne . Uniaxial pressure dependencies of opposite sign, $-1.3$~K/GPa and $0.3$~K/GPa, are extracted for the out-of-plane and in-plane directions at \TNOne. For $T^*$ we obtain ${\partial}T^*/{\partial}p_{\mathrm{c}} = 1.4$~K/GPa. In particular we elucidate thermodynamic properties of the recently discovered skyrmion lattice phase and show that it is strongly enhanced by uniaxial pressure. 
\end{abstract}

	 \maketitle

\section{Introduction}
Ternary intermetallic compounds of the type \ce{R2TX3} ($R$ = rare earth, $T$ = transition metal, $X$ = element of main groups III to V)~\cite{Hoffmann-Poettgen2001, Pan2013} have been investigated extensively over the past decades, due to their variety of intriguing electronic properties ranging from superconductivity~\cite{Majumdar2001}, giant magnetoresistance (GMR)~\cite{Majumdar2000, Majumdar2001GMR, Paulose2003}, ferromagnetism~\cite{Cao2010} and incommensurate spin structures~\cite{Kurumaji2019, Hirschberger2020PRB}, phenomena related to Kondo physics and heavy fermions~\cite{Majumdar1999, Saha2000, Majumdar2002}, to non-Fermi-liquid~\cite{Majumdar1999} and spin-glass behavior~\cite{Kaczorowski1993, Tien1997, Li1999, Nimori2003}. This is particularly evident in the title material \GPS\ where a skyrmion lattice phase featuring giant topological Hall and Nernst effect was discovered recently.~\cite{Saha1999,Kurumaji2019,Hirschberger2020PRL}

Most members of the \ce{R2PdSi3} family of ternary silicides crystallize in a highly symmetric \ce{AlB2}-derived hexagonal structure (space group P6/\textit{mmm}) with triangular lattice layers of $R^{3+}$ magnetic sites sandwiching honeycomb nets of Pd/Si sites. While the Pd and Si ions were originally believed to be distributed statistically~\cite{Szytula1999}, an X-ray and neutron diffraction study by Tang et al.~showed for \ce{Ho2PdSi3} that these ions actually order into a superstructure along both in- and out-of-plane directions,
while the overall centrosymmetry of the structure is retained~\cite{Tang2011}. 
This leads to two nonequivalent sites for the $R^{3+}$ ions, which has been shown to affect the magnetism in an applied magnetic field for \ce{Er2PdSi3}~\cite{Tang2010}.
While no structural phase transition has been detected for R = Gd, Tb, Dy, Ho, Er and Tm, most \ce{R2PdSi3} compounds show long-range magnetic order at low temperatures~\cite{Frontzek2009, Smidman2019, Mukherjee2011}.

These various ordering phenomena are driven by a delicate interplay of indirect exchange coupling mediated by the conduction electrons, i.e., the Rudermann-Kittel-Kasuya-Yosida (RKKY) interaction, spin-orbit coupling and the influence of crystal field (CF) effects.
The Gd$^{3+}$ ions in \GPS, however, with a half-filled $4f$ shell, have vanishing orbital momentum (\textbf{J} $\approx$ \textbf{S} $= 7/2$) and are not influenced by crystal field effects. Magnetic order, therefore, arises from the RKKY interaction and dipole-dipole interactions.
Gd$_2$PdSi$_3$ exhibits two successive phase transitions around \TN~=~21~K~\cite{Hirschberger2020PRB} and was found to exhibit a skyrmion lattice (SkL) phase of Bloch-type skyrmions in low magnetic fields applied along the $c$-axis~\cite{Kurumaji2019}. A number of incommensurate spin structures both in zero-field as well as in higher applied magnetic fields have been identified~\cite{Hirschberger2020PRB} and the phase diagram in fields up to 9~T has been established through resistance and magnetization measurements as well as resonant X-ray scattering~\cite{Frontzek2009, Kurumaji2019, Hirschberger2020PRB, Zhang2020}.
Single crystal X-ray and neutron diffraction measurements yielded lattice parameters at 300~K (2~K) of $a$~=~4.079~\r{A} (4.066~\r{A}) and $c$~=~4.098~\r{A} (4.091~\r{A}), i.e. $\Delta a/a$~=~$3.2\cdot 10^{-3}$ and $\Delta c/c$~=~$1.7\cdot 10^{-3}$.~\cite{Tang2011}$^{,}$\cite{FN4}

Except for these measurements of the lattice parameters, however, there is at present no study on magnetoelastic effects in \GPS. Therefore, with a particular focus on the skyrmion lattice phase, we report detailed dilatometric studies of \GPS~in a wide range of temperatures and magnetic fields. Our thermal expansion and magnetostriction data show pronounced magnetoelastic coupling and field effects extending up to temperatures of 150~K and above. Moreover, we uncover yet unreported phases and an anomaly in zero-field which appears well below the N\'{e}el transitions at \TNOne~=~22.3(5)~K and \TNTwo~=~19.7(5)~K, thereby evidencing competing interactions already in zero-field. We update the magnetic phase diagram for $B\parallel c$-axis, present for the first time the phase diagram for $B \parallel a$*-axis, and discuss in detail the thermodynamic properties for $B\parallel c$. Our results in particular elucidate the skyrmion lattice phase and we show that it is enhanced by uniaxial pressure.

\section{Experimental Methods}
Single crystals of \GPS\ have been grown by the optical floating-zone method as reported in Ref.~[\onlinecite{Mazilu2006, Xu2011}] and were previously studied by AC susceptibility, neutron diffraction~\cite{Frontzek2009}, and angle-resolved photoemission spectroscopy~\cite{Inosov2009}.
The magnetization was studied in the temperature regime from 1.8~K to 300~K in magnetic fields up to 7~T in a Magnetic Properties Measurement System (MPMS3, Quantum Design) and up to 14~T in a Physical Properties Measurement System (PPMS, Quantum Design) using the Vibrating Sample Magnetometry (VSM) option. Specific heat measurements were performed on a PPMS-14 using a relaxation method on single crystals of $m=20.79$~mg (2~K-300~K) and 13.49~mg (0.15~K-3~K). High-resolution dilatometry measurements were performed by means of a three-terminal high-resolution capacitance dilatometer in a home-built setup placed inside a Variable Temperature Insert (VTI) of an Oxford magnet system~\cite{Kuechler2012,Werner2017}. With this dilatometer, the relative length changes $dL_i/L_i$ along the crystallographic $c$ and $a$* directions, respectively, were studied on an oriented cuboid-shaped single crystal of dimensions $2.480 \times 1.300 \times 1.459~$mm$^{3}$. Measurements were performed at temperatures between 2~K and 300~K in magnetic fields up to 15~T, applied along the direction of the measured length changes, and the linear thermal expansion coefficients $\alpha_i=1/L_i\cdot dL_i(T)/dT$ were derived. In addition, the field-induced length changes $dL_i(B_i)$ were measured at various fixed temperatures between 1.7~K and 200~K in magnetic fields up to 15~T. The longitudinal magnetostriction coefficient $\lambda_i~=~1/L_i~\cdot~dL_i(B_i)/dB_i$ was derived from $dL_i(B_i)$.

\section{Experimental Results}

\subsection{Evolution of magnetic order at $B=0$}

\begin{figure}[htbp]
	\center{\includegraphics [width=0.9\columnwidth,clip]{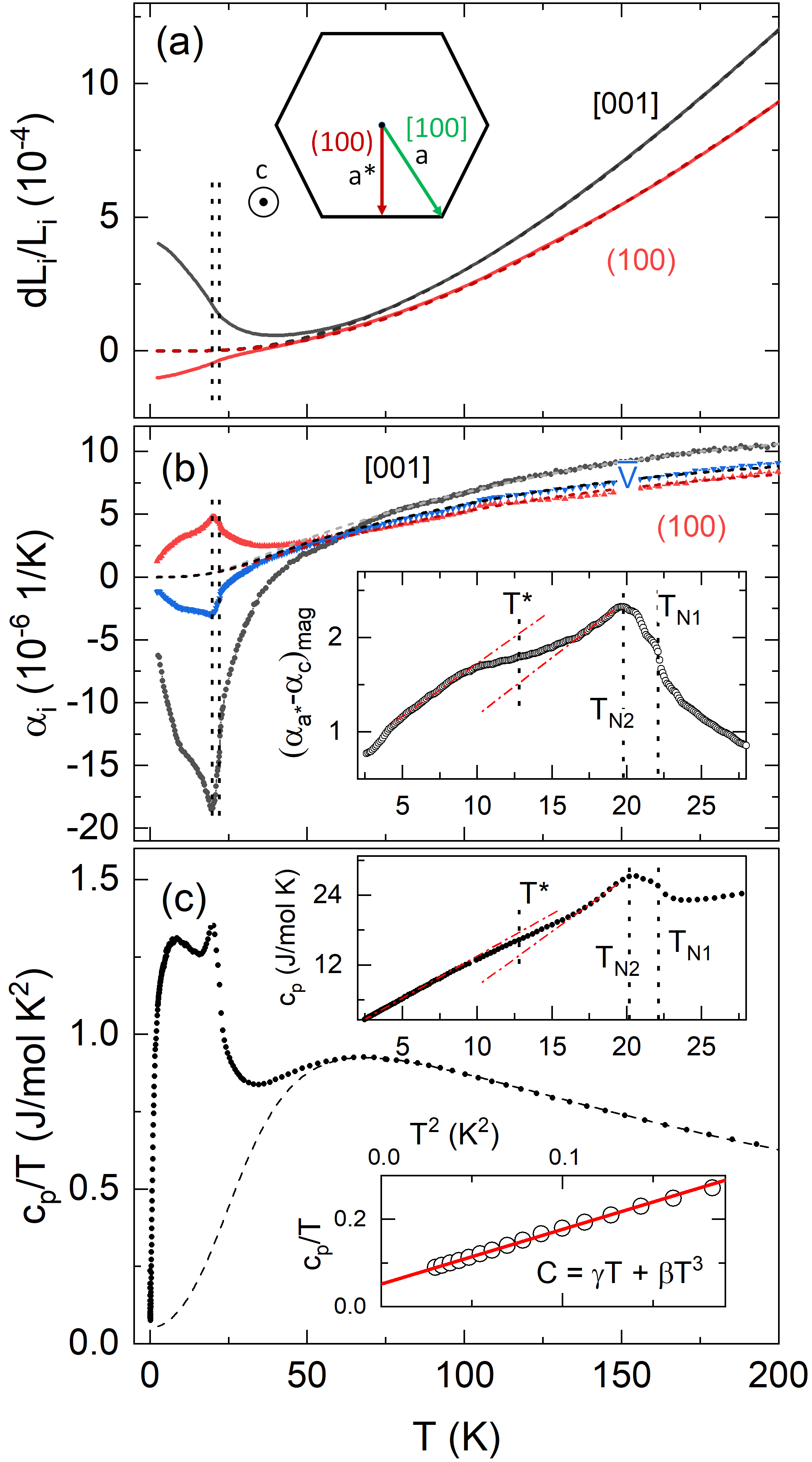}}
	\caption[] {\label{alpha_0T}(a) Relative length changes $dL_i/L_i$ along the $c$ and $a$*-crystallographic directions. Directions w.r.t. the Gd lattice are indicated in the inset. (b) Associated linear thermal expansion coefficients $\alpha_i$ as well as the calculated 1/3 volume expansion $\alpha_{\mathrm{V}}$. The inset displays the difference $\alpha_{\mathrm{a^*,mag}} - \alpha_{\mathrm{c,mag}}$ of the background corrected thermal expansion coefficients $\alpha_{\mathrm{i,mag}} = \alpha_i- \alpha_{\mathrm{ph}}$. Vertical dashed lines indicate two phase transitions and the dashed-dotted line (inset) is a guide to the eye highlighting a feature around $T^*$. (c) Specific heat \cpt~(black markers). Inset: Low-temperature regime plotted as \cpt~vs. $T^2$. The solid red line shows a fit with \cp~$=\gamma T+\beta T^3$. Dashed curves in all graphs mark the non-magnetic background as explained in the text.}
\end{figure}

Uniaxial thermal expansion and specific heat show pronounced anomalies around 20~K which are associated with the onset of long-range magnetic order (Fig.~\ref{alpha_0T}). Close inspection of the anomalies indicates the proximity of not only one, but two phase transitions around \TN. While the anomaly at \TNOne~= 22.3(5)~K is seen as a jump in $\alpha_i$, the anomaly at \TNTwo~= 19.7(5)~K (20.3(5)~K for the $a$*-axis) is expressed as a peak. This observation confirms the results by Hirschberger et al.~of two consecutive phase transitions in zero field~\cite{Hirschberger2020PRB}.
Beyond these two known transitions, however, our data display a third anomaly indicative of a phase transition which has not been reported for single crystals of \GPS. This anomaly, marked by $T^*\approx 13$~K in the inset of Fig.~\ref{alpha_0T}(b), is visible as a broad jump for both directions in \ali\ which extends between 10~K and 15~K and will be discussed in more detail below. 

The specific heat data show the three observed features at \TNOne, \TNTwo~and $T^*$, too (Fig.~\ref{alpha_0T}(c)). The shape of the anomalies in \cp\ is analogous to their shape in \ain.
At very low temperatures below about 400~mK a quasi-linear behavior of \cpt\ vs. $T^2$ is found (see the inset of Fig.~\ref{alpha_0T}(c)) reminding of similar observations in \LPS\ and \CPS .~\cite{Cao2013,Saha2000}  In this temperature regime, the data are described well by a linear and a cubic term, i.e., \cpt\ $ = \gamma +\beta T^2$. The quasi-linear term is described by an effective Sommerfeld coefficient $\gamma = 52(5)$~mJ/(mol K$^2$). This parameter is in between the values obtained for \LPS\ (6.9~mJ/(mol K$^2$)) and \CPS\ (108~mJ/(mol K$^2$)), the latter being discussed as evidence of heavy-fermion behaviour.~\cite{Saha2000} Whereas phonons can be neglected in this temperature regime, the coefficient $\beta = 1.25(3)$~J/(mol K$^4$) is rather large and reflects the contribution of low-energy antiferromagnetic excitations.

The dashed lines in Fig.~1 show the phononic and electronic contributions to the relative length changes, thermal expansion coefficients, and specific heat. In order to obtain these contributions, the specific heat of the non-magnetic analog \LPS~as reported by Cao et al.~\cite{Cao2013} was fitted by phononic Debye and Einstein terms, as well as an electronic term, according to 

\begin{equation}\label{eq:DebyeEinstein_cp}
    c_p^{el,ph}= \gamma T + n_{\mathrm{D}}D\left(\frac{T}{\Theta_{D}}\right)+n_{\mathrm{E}}E\left(\frac{T}{\Theta_{E}}\right)
\end{equation}

where $\gamma$ is the Sommerfeld coefficient, $n_{\mathrm{D}}$ and $n_{\mathrm{E}}$ are constants, $D(T/\Theta_{\mathrm{D}})$ and $E(T/\Theta_{\mathrm{E}})$ are the Debye and Einstein functions with the Debye and Einstein temperatures $\Theta_{\mathrm{D}}$ and $\Theta_{\mathrm{E}}$. The fit to the \LPS~specific heat data yields $\Theta_{\mathrm{D}}$ = 213~K, $\Theta_{\mathrm{E}}$ = 454~K, with $n_{\mathrm{D}} = 3.69$ and $n_{\mathrm{E}} = 1.98$. $\gamma$ was fixed to the value reported by Cao et al.~of 6.93~mJ/(mol K$^2$). Compared to $\Theta_{\mathrm{D}} = 191$~K by Cao et al., extracted from the low temperature regime, our value is slightly larger.

Scaling the Debye and Einstein temperatures by the different masses of Lu and Gd we obtain a scaling factor~\cite{Tari2003} of $\Theta_{\mathrm{D, LPS}}/\Theta_{\mathrm{D, GPS}} = 0.962$. The specific heat and thermal expansion of \GPS~were thus fitted with fixed $\Theta_{\mathrm{D}} = 222$~K $= \Theta_{\mathrm{D, LPS}}/0.962$ and correspondingly $\Theta_{\mathrm{E}} = 471$~K.
For the fit to the specific heat, $\gamma = 52$~mJ/(mol K$^2$) was also fixed.
For the thermal expansion the electronic contribution was negligibly small and therefore omitted, i.e. it was fitted by
\begin{equation}\label{eq:DebyeEinstein_alpha}
    \alpha^{ph}= n_{\mathrm{D}}D\left(\frac{T}{\Theta_{D}}\right)+n_{\mathrm{E}}E\left(\frac{T}{\Theta_{E}}\right)
\end{equation}
with parameters $n_{\mathrm{D}}$ and $n_{\mathrm{E}}$.
The phononic contributions to $dL_i/L_i$ in Fig.~1(a) were obtained by integrating the background obtained for the respective \ali. 

Subtracting the electronic and phononic backgrounds from the specific heat and  thermal expansion coefficients yields their respective magnetic contributions which extend up to about 60~K. This agrees with the temperature regime where the magnetization exhibits a non-linear field dependence up to 7~T (see Fig.~\ref{SIchi}). From \cptmag~the changes in magnetic entropy, $S_{\mathrm{mag}}$, above 150~mK are calculated. We obtain a constant $\Delta S_{\mathrm{mag}}(T>150~\mathrm{mK})= 31.3$~\jmk\ above 60~K, which is 90\% of the full expected magnetic entropy of 2$R\ln{8} = 34.6$~\jmk, where $R$ is the universal gas constant.

Returning to the thermal expansion data, we see that the anomalies in the thermal expansion coefficients, at \TNOne\ and \TNTwo, are of opposite sign for the $c$- and $a$*-axis, indicating opposite pressure dependencies ${\partial}T_{\mathrm{N}i}/{\partial}p_{\mathrm{c}} <$~0 and ${\partial}T_{\mathrm{N}i}/{\partial}p_{\mathrm{a^*}} >$~0. 
The volume thermal expansion also indicates a negative hydrostatic pressure dependence ${\partial}T_{\mathrm{N}i}/{\partial}p <$~0 for both antiferromagnetic transitions.

The Gr\"{u}neisen ratio of the thermal expansion coefficient and the specific heat is a valuable quantity to determine the relevant energy scales driving the system and to quantify its pressure dependencies. In the presence of one dominant energy scale $\epsilon$, this ratio is independent of temperature and enables the determination of the pressure dependence of $\epsilon$, i.e.~\cite{Gegenwart2016Grueneisen,Klingeler2006pressure}, 

\begin{equation}
\Gamma_{i} = \frac{\alpha_{i}}{c_{\mathrm{p}}} = \frac{1}{TV_{\mathrm{m}}}\frac{\partial S/\partial p_i}{\partial S/\partial T}= \frac{1}{V_{\mathrm{m}}}\frac{\partial \ln \epsilon}{{\partial}p_{i}}.   \label{eq:gruen}
\end{equation}

\begin{figure}[htbp]
	\center{\includegraphics [width=0.9\columnwidth,clip]{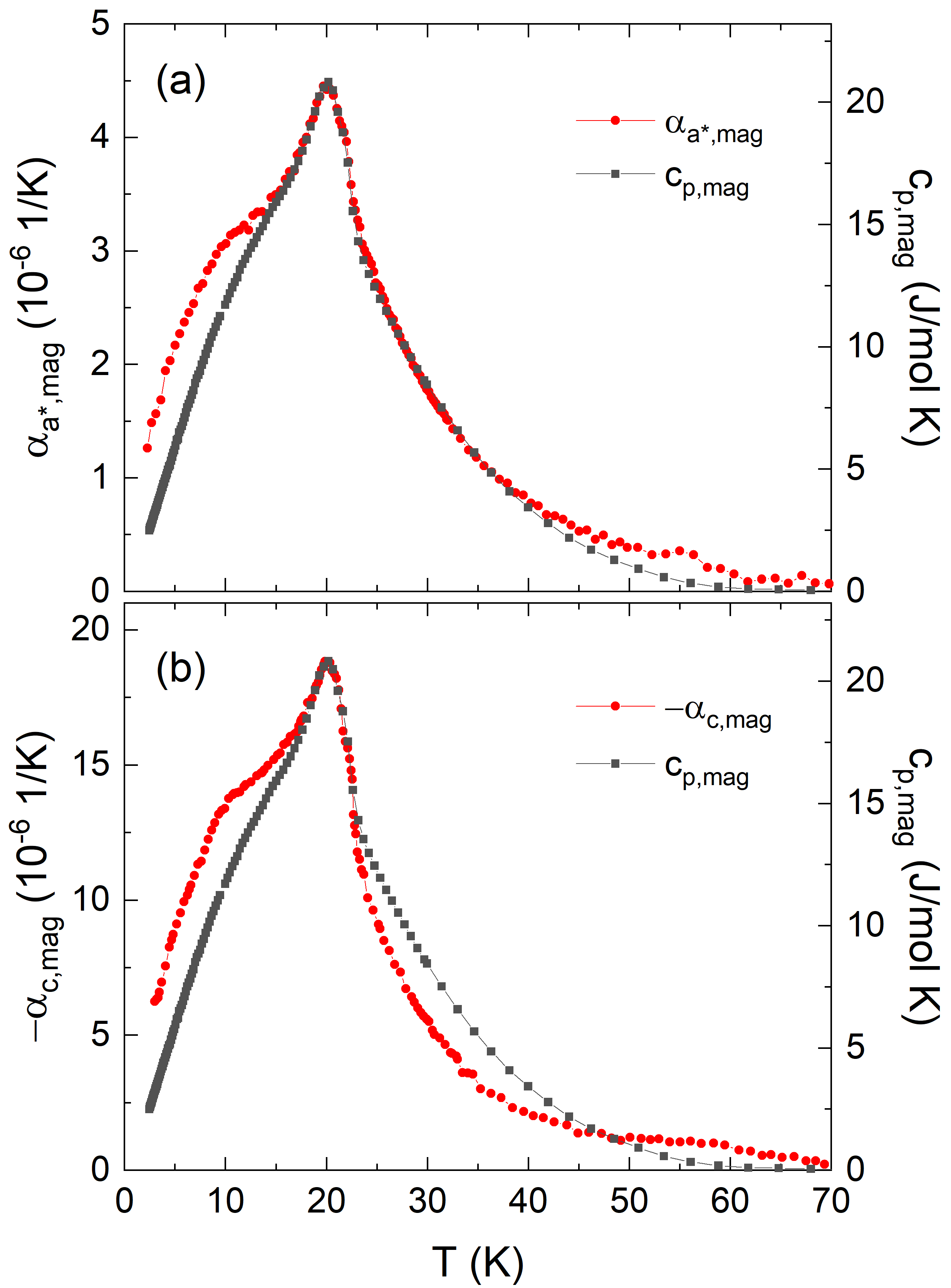}}
	\caption[a] {\label{a-vs-Cp} Magnetic contributions to the thermal expansion coefficient (left axis) and specific heat (right axis) for (a) the $a$*-axis and (b) the $c$-axis after subtracting phononic and electronic contributions as described in the text.~\cite{FN1}}
\end{figure}

Here, $V_{\mathrm{m}}$ is the molar volume and the index $i$ indicates a linear direction or the volume. At \TN , Eq.~\ref{eq:gruen} converts to $\Gamma$ = (\TN $V_m$)$^{-1}\cdot \partial $\TN /$\partial p$. 
Comparing the magnetic contributions \aimag\ and \cpmag\ hence allows to identify temperature regimes where the Gr\"{u}neisen relation implies only one dominant energy scale while appropriate scaling enables to read off the respective parameter $\Gamma_{\mathrm{i,mag}}$. As shown in Fig.~\ref{a-vs-Cp}, the overall behavior of \aimag\ and \cpmag\ is similar except for a distinct jump in \aimag\ at $\sim 12$~K which is much less pronounced in the magnetic specific heat. In both cases, magnetic contributions start to evolve around 60~K.

Despite the overall similar behavior, there are differences at higher temperatures, too. While the $a$*-axis shows a nearly perfect overlap between \aimag\ and \cpmag\ down to 14~K as shown in Fig.~\ref{a-vs-Cp}(a), we only observe a very good agreement around \TNOne~and \TNTwo~for the $c$-axis, in a range from 17~K to about 23~K. We also note that below $\sim$14~K our results indicate the failure of Gr\"{u}neisen scaling rather than the presence of just a different scaling parameter. 

Our data, however, clearly imply the presence of a single dominant energy scale at and around the magnetic ordering temperatures \TNOne~and \TNTwo. The obtained Gr\"{u}neisen parameters amount to $\Gamma_{\mathrm{c,mag}} = -91(13)\cdot 10^{-8}$~mol/J and $\Gamma_{\mathrm{a^*,mag}}= 22(3)\cdot 10^{-8}$~mol/J. From these values, moderate pressure dependencies are derived, i.e., we obtain negative pressure dependencies $\partial T_{\mathrm{N1}}/\partial p_{\mathrm{c}} = -1.3(2)$~K/GPa and $\partial T_{\mathrm{N2}} /\partial p_{\mathrm{c}} = -1.4(2)$~K/GPa for uniaxial pressure applied along the $c$-axis. The uniaxial pressure dependencies for $p\parallel a$*-axis are positive and more than a factor of four smaller, i.e., 0.31(5)~K/GPa for \TNOne~and 0.34(5)~K/GPa for \TNTwo.

While our data hence evidence that the ordering phenomena at \TNOne\ and \TNTwo\ are governed by the same energy scale, an additional energy scale becomes relevant upon further cooling, around $T^*$, as proven by the failure of Gr\"{u}neisen scaling (cf. Fig.~\ref{a-vs-Cp}). Closer inspection of the associated anomalies implies not only a broad jump-like increase in the thermal expansion coefficients but also a less pronounced anomaly in \cp\ which is visible much more clearly in the \cp /$T$ data in Fig.~\ref{alpha_0T}c. In an attempt to deduce the anomaly size associated with the respective features we obtain $\Delta c_{\mathrm{p}}^*\approx 2.7(5)$~\jmk , $\Delta\alpha_{\mathrm{V}}^*\approx 1.8\cdot 10^{-6}$/K, $\Delta\alpha_{\mathrm{c}}^*\approx 4.1(6)\cdot 10^{-6}$/K, $\Delta\alpha_{\mathrm{a^*}}^*\approx -1.0(3)\cdot 10^{-6}$/K. The changes in magnetization around T$^*$ are very small for both axes and could not be seen in the isothermal magnetization $M(B)$. However, temperature sweeps of the magnetization in static field evidence a jump in ${\partial}\chi/{\partial}T$, which is visible for $B \geq 0.2$~T (0.25~T) for $B\parallel c$ ($B\parallel a$*) (see SI, Fig.~\ref{SI_T-star}(a)). At 0.2~T the jump height amounts to 5.6(1.4)$\cdot 10^{-3} \mu_{\mathrm{B}}$/(f.u. K).
Further values are listed in Tab.~\ref{tab:T-star}.

\subsection{Thermal Expansion at B~$\neq$~0 and Magnetostriction}

\begin{figure}[htbp]
	\center{\includegraphics [width=0.9\columnwidth,clip]{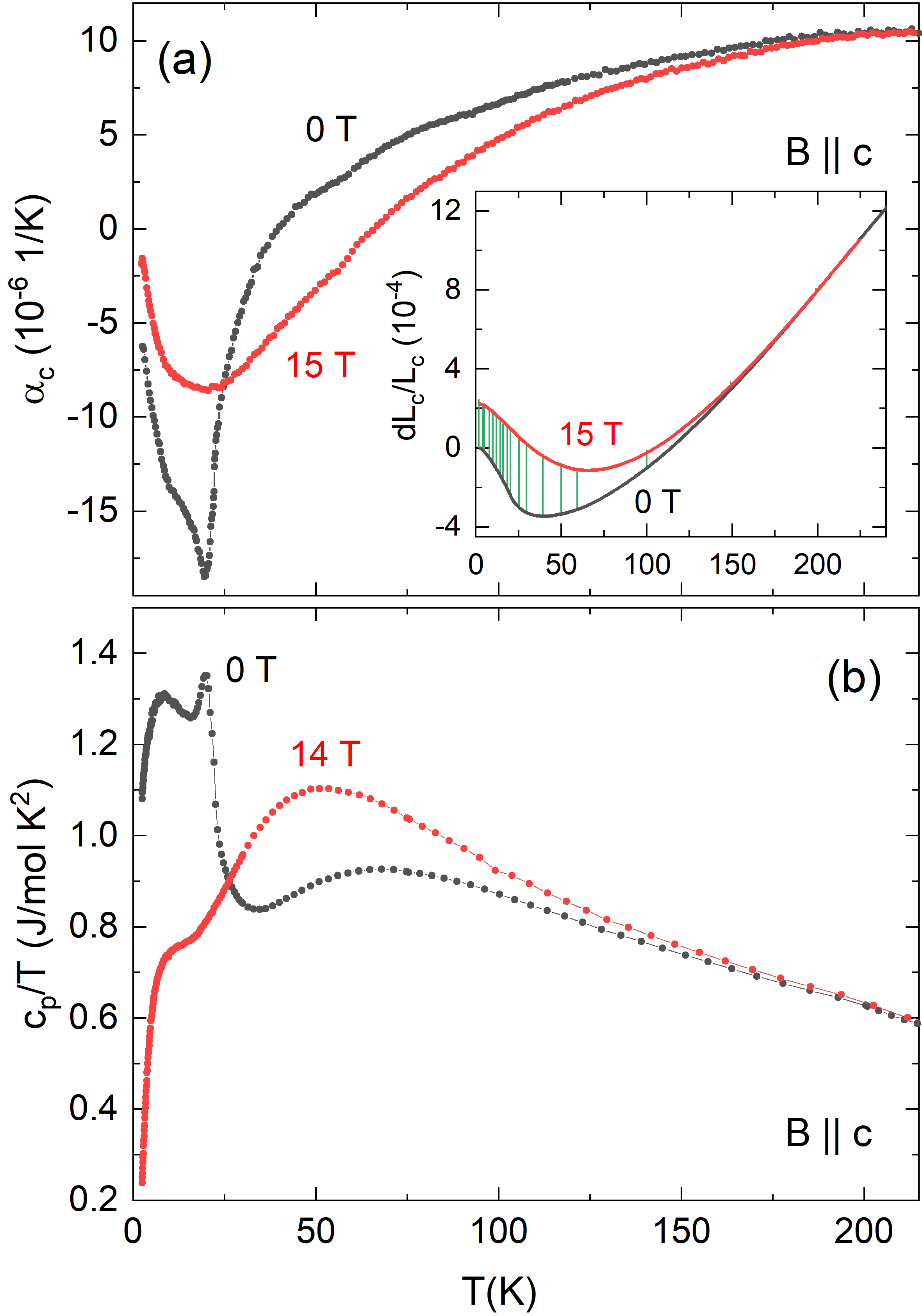}}
	\caption[a] {\label{Alpha_0T-and-15T} The effect of high fields on (a) the thermal expansion coefficient ($B$~=~15~T) and (b) the specific heat \cpt~($B$~=~14~T) as compared to zero-field measurements. The inset in (a) shows the relative length changes. Vertical green bars indicate magnetostriction data from 0~T to 15~T at several temperatures.}
\end{figure}

The effect of high magnetic fields on the thermal expansion and specific heat is shown in Fig.~\ref{Alpha_0T-and-15T}. A number of observations can be made: (1) The sharp features indicating phase transitions are absent at $B=15$~T. (2) Significant entropy is shifted to higher temperatures and, at $B=15$~T, significant field effects are visible at least up to 150~K in all shown quantities, in particular for $\alpha_{\mathrm{c}}$ even up to about 200~K. (3) Magnetostriction from 0~T to 15~T is positive (negative) for the $c$-axis ($a$*-axis), and (4) the temperature region of negative thermal expansion of the $c$-axis extends up to about 65~K at 15~T, compared to 38~K in zero-field. Note, that the magnetostriction data fully agree to the thermal expansion data at $B\neq 0$~T as shown by the (green) vertical lines in the inset of Fig.~\ref{Alpha_0T-and-15T}).

\begin{figure}[ht]
	\center{\includegraphics [width=1\columnwidth,clip]{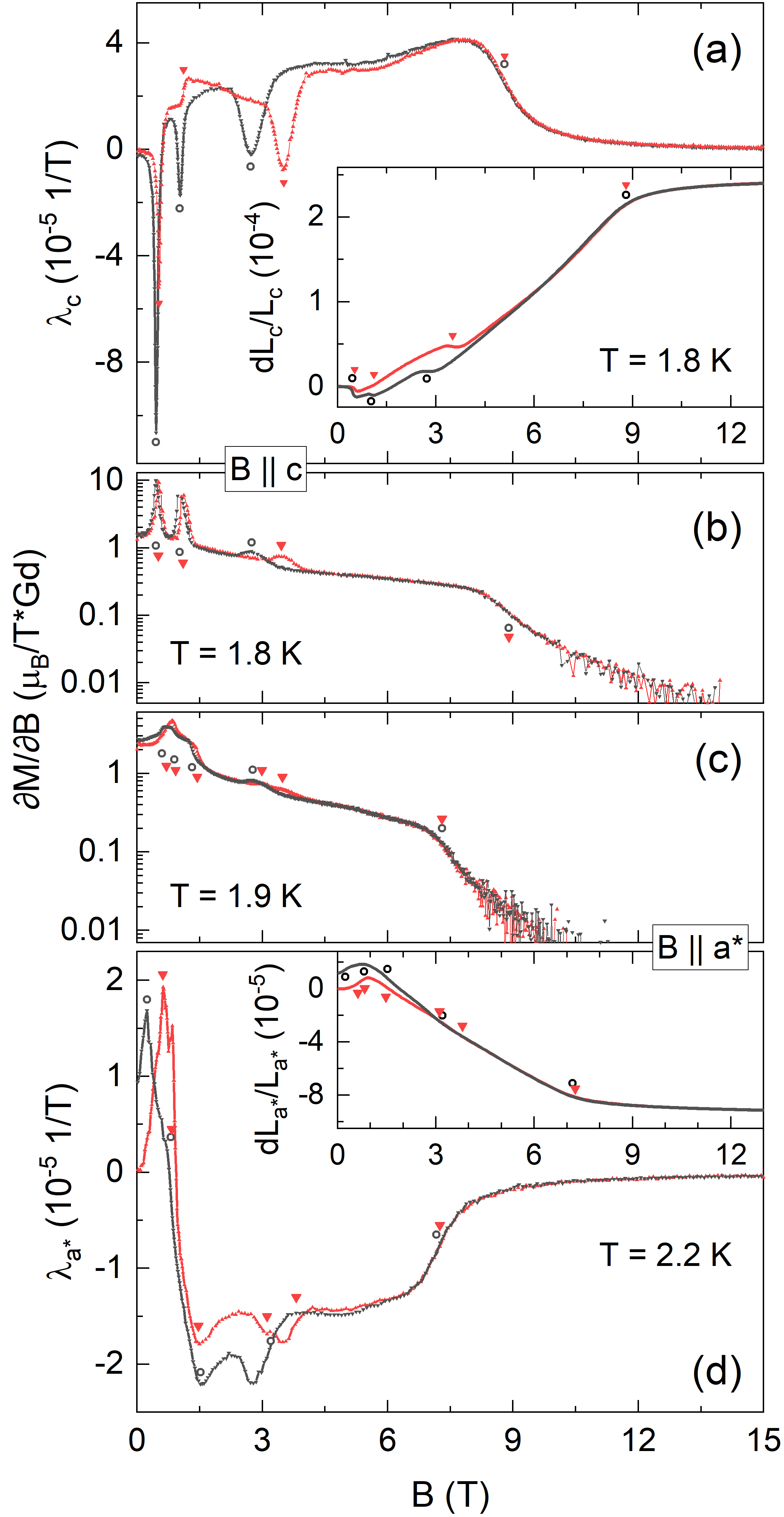}}
	\caption[] {\label{MS_Low-T} Magnetostriction coefficients $\lambda_i$ ($i$ = $c$, $a$*) and isothermal magnetic susceptibility, $\partial M/\partial B$, at temperatures around 2~K, for $B\parallel c$ (a-b) and $B\parallel a$* (c-d). Note the logarithmic scale in (b) and (c). Insets in (a) and (d) show the relative length changes $dL_\mathrm{i}(B)/L_\mathrm{i}$. Triangles and empty circles mark anomalies in (a-d). Red markers and lines represent up-sweeps, black ones down-sweeps. }
\end{figure}

Magnetostriction data at $T\approx 2$~K shown in Fig.~\ref{MS_Low-T}(a) and (d) further confirm strong magnetoelastic coupling and in addition clearly show the field-induced phase transitions. For comparison the isothermal magnetic susceptibility $\chi (B) =\partial M(B)/\partial B$ is also presented on the same field scale for both directions (Fig.~\ref{MS_Low-T}(b-c)). Considering the data for $B\parallel c$, four anomalies can be identified (Fig.~\ref{MS_Low-T}(a)): Up to 3.5 T, there are two sharp peaks in $\lambda_{\mathrm{c}}$ signalling jumps in $dL_{\mathrm{c}}(B)$ with only small field-hysteresis, followed by a broad peak with a large hysteresis of $\sim$0.8~T. The size of the anomalies for up- and down-sweep differs strongly.
All three anomalies indicate discontinuous phase transitions. Corresponding anomalies and hystereses are also visible in the magnetic susceptibility. In addition, there is a broad downward jump in $\lambda_{\mathrm{c}}$ at around 9~T, above which magnetostriction becomes virtually zero which is also reflected by small $\chi$, i.e., rather full alignment of magnetic moments in field (please note the logarithmic scale in Fig.~\ref{MS_Low-T}(b) and (c)). The overall region where hysteresis is visible extends from about 6~T down to the lowest fields (see the inset in Fig.~\ref{MS_Low-T}(a)) but no remanent magnetostriction is visible which would indicate irreversible changes in the sample, e.g., through domain effects. Four features are also visible in $\lambda_{\mathrm{a^*}}$ for $B \parallel a$*  (Fig.~\ref{MS_Low-T}(d)). These anomalies are smaller in magnitude, much broader and less well-defined than for the $c$-axis. Similar to the findings for $\lambda_{\mathrm{c}} (B \parallel c)$, there is a jump at higher fields, at about 7.3~T, but here of opposite sign. Again, it signals a continuous transition to the saturated phase of vanishing magnetostriction. In contrast to $B \parallel c$, the magnetostriction measurements $dL_{\mathrm{a^*}}$($B \parallel a$*) feature pronounced remanent magnetostriction below 5~K, i.e., non-zero overall length changes after sweeping the field from 0~T to 15~T and back to 0~T. At 1.8~K, this amounts to $(\Delta L/L)_{\mathrm{rem}} = 1.4\cdot 10^{-5}$. We attribute this observation to the irreversible effects of structural or magnetic domains as seen, e.g., in \ce{CoCl2}~\cite{Kalita2000}, \ce{NiCl2}~\cite{Kalita2002} and \ce{NiTiO3}~\cite{Dey2021}. Such irreversible domain effects seem to be absent in the measurements along the $c$-axis. The transition between a multidomain and single domain state may thus be fully reversible for $B\parallel c$. We conclude that hysteresis found for $B\parallel a$* below 3.5~T is both due to the discontinuous nature of the phase transitions and domain effects.

Both data sets, hence, imply a series of four phase transitions in magnetic field, at $T\approx 2$~K, which is also corroborated by magnetization studies (also see Fig.~\ref{SI_Magnetometry}) and agrees to the recently published phase diagram for $B\parallel c$.~\cite{Hirschberger2020PRB} Following the notations in Refs.~[\onlinecite{Hirschberger2020PRB, Kurumaji2019}] for the phases appearing for $B\parallel c$, we label the respective phases as IC-1, A, IC-2, DP, and field induced ferromagnetic (fiFM) phase, with IC-1/IC-2 being characterized by incommensurate spin configurations, A by the formation of a skyrmion lattice (SkL), and DP by the depinning of the direction of magnetic moments (see also the phase diagram in Fig.~\ref{PD_All}). We note, however, that while the magnetostriction data evidence field-driven structural changes, domain effects may obscure the actual phase transitions up to the field and temperature regions at which a single domain state is achieved. In particular, broad peaks in the magnetostriction coefficients as seen in $\lambda_{\mathrm{a^*}}$ (Fig.~\ref{MS_Low-T}(b), \ref{SI_MS_Low}(e) and (f)) do not necessarily indicate the actual phase boundaries, but the peak positions may differ from those found in the magnetization studies, as shown by a phenomenological model by Kalita et al.~\cite{Kalita2000}. Therefore, for the further thermodynamic analysis of the phase boundaries as well as the construction of the phase diagram, for $B \parallel a$* we will only consider anomalies in the magnetostriction which can directly be linked to anomalies in isothermal magnetization.

\begin{figure}[htbp]
	\center{\includegraphics [width=0.9\columnwidth,clip]{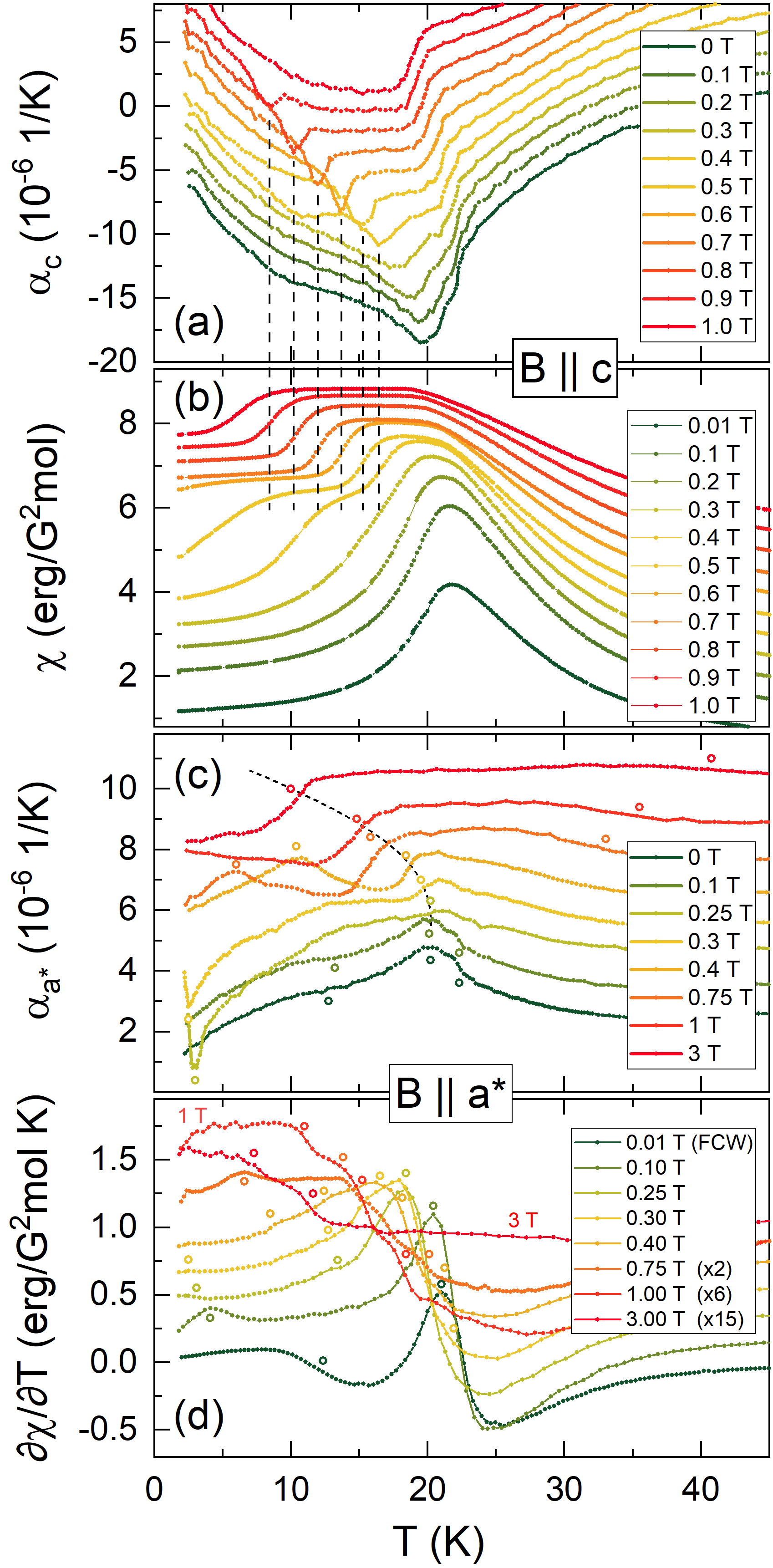}}
	\caption[] {\label{TE-and-M-of-T} Thermal expansion coefficients (a,c) and static magnetic susceptibility $\chi = M/B$ (b), respectively its derivative $\partial \chi/\partial T$ (d), for (a-b) $B\parallel c$ and (c-d) $B\parallel a$*. Curves are offset vertically by (a) 1.6$\cdot 10^{-6}$/K, (b) 0.5~erg/(G$^2$~mol), (c) 1$\cdot 10^{-6}$/K, and (d) 0.2~erg/(G$^2$~mol~K), respectively, for better visibility. 0.75~T to 3~T data in (d) are multiplied by a constant as indicated in the legend and 3~T data is offset by 1.5~erg/(G$^2$~mol~K) instead of 1.4~erg/(G$^2$~mol~K). "FCW" indicates a field-cooled-warming measurement.~\cite{FCW} Empty circles mark temperature positions of the anomalies as extracted for the phase diagram.}
\end{figure}

In order to obtain the phase boundaries, we have performed thorough magnetostriction and isothermal magnetization studies at various fixed temperatures as well as corresponding temperature sweeps at fixed magnetic field (see Figs.~\ref{SI_MS_Low},~\ref{SI_MS_High},~\ref{SI_Magnetometry}).~\cite{FN2} This is demonstrated in Fig.~\ref{TE-and-M-of-T}, where the thermal expansion coefficients in low fields up to 3~T and the corresponding magnetization data are presented. For $B\parallel c$ (Fig.~\ref{TE-and-M-of-T}(a) and (b)), the evolution of two different phase boundaries can be traced straightforwardly. Specifically,  applying small fields yields a suppression of \TNTwo\ while the jump at \TNOne\ becomes more distinguished. In increasing field \TNOne\ is also suppressed to lower temperature, but to a smaller extent (Fig.~\ref{TE-and-M-of-T}a). Above 0.3~T the peak assigned to \TNTwo\ at zero-field changes its shape, signaling the transition from the previously reported skyrmion lattice A phase to the IC-2 phase. It is associated with a jump \dchicA\ in the static susceptibility from one $T$-independent value to another (Fig.~\ref{TE-and-M-of-T}(b)). At 0.4~T, a second feature corresponding to the transition from the A phase to the IC-1 phase is visible in both \ac\ and $\chi_{\mathrm{c}}$, while above 0.9~T (1~T for $\chi_{\mathrm{c}}$) all features below \TNOne\ are gone. Quantitatively, \dchicA\ gradually decreases from 0.91~$\mu_{\mathrm{B}}$/Gd$^{3+}$ at 0.9~T and 8.6~K to 0.36~$\mu_{\mathrm{B}}$/Gd$^{3+}$ at 0.4~T and 16.4~K. 
The pronounced jump in \ac\ at \TNOne\ corresponds to a kink in the static susceptibility, i.e., a jump in its derivative (see SI, Fig.~\ref{SI_Magnetometry}).

As mentioned before, anomalies seen for measurements along the $a$*-axis are in general much weaker and less well-defined than for the $c$-axis. Furthermore, the evolution of anomalies in the thermal expansion and the static susceptibility along the $a$*-axis is even more complex than for $B\parallel c$ (Fig.~\ref{TE-and-M-of-T}(c) and (d)).
In zero-field, the anomaly at $T^*$ is also visible (see Fig.~\ref{alpha_0T}, as well as SI, Fig.~\ref{SI_T-star}(d) and can be traced up to 0.4~T in ${\partial}\chi/{\partial}T$.
Also, a jump in \ain~evolving from \TNOne\ can be followed to lower temperatures for increasing fields, corresponding to a jump in $\partial \chi/\partial T$. 
Above 0.6~T this jump splits into two jumps, uncovering an additional phase between the IC-2 and fiFM phases, while we only see one broad jump in \ain.

\section{Discussion}

\begin{figure}[tb]
	\center{\includegraphics [width=1\columnwidth,clip]{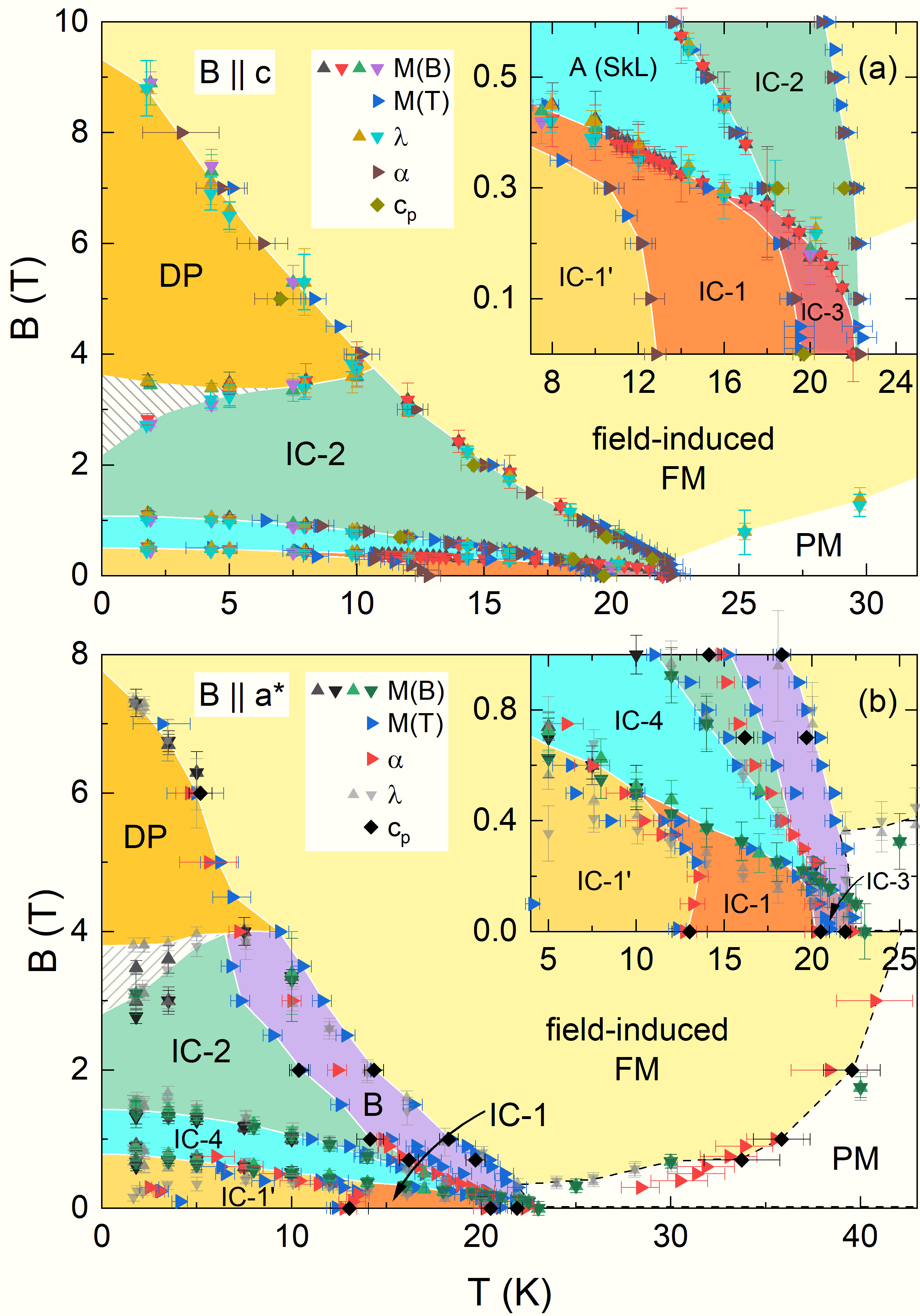}}
	\caption[] {\label{PD_All} Phase diagrams for (a) $B \parallel c$ and (b) $B \parallel a$* constructed from different experimental techniques as indicated in the legends. The shaded areas show two strong hysteresis regimes. The abbreviations for the phases are: paramagnetic (PM), incommensurate magnetic orders (IC-x), antiferromagnetic (A,B), antiferromagnetic skyrmion lattice (SkL), and depinned phase (DP).}
\end{figure}

\begin{figure*}[htbp]
	\center{\includegraphics [width=0.7\columnwidth,clip]{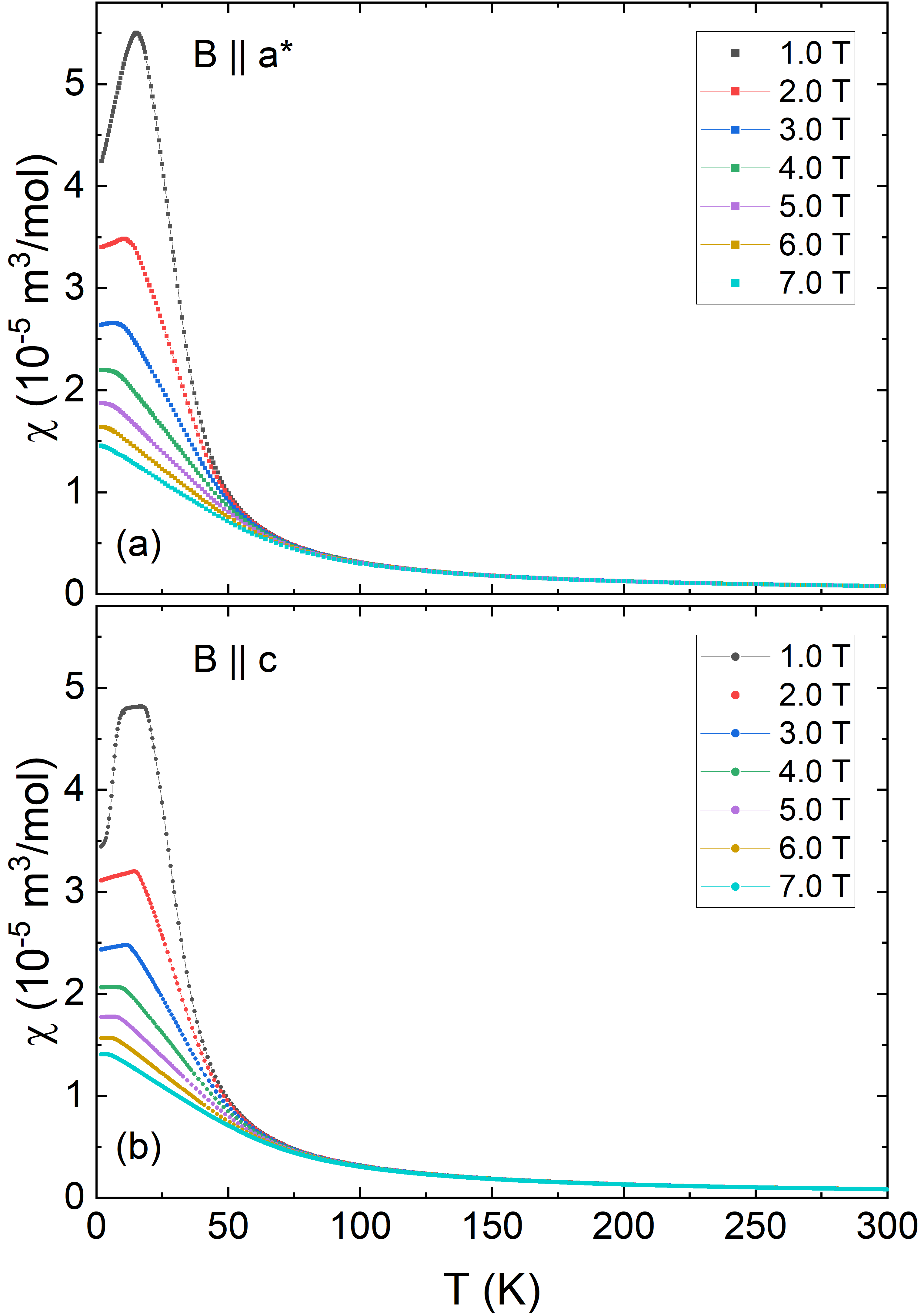}}
	\caption[] {\label{SIchi}Static magnetic susceptibility $\chi = M/B$ for (a) $B\parallel a$* and (b) $B \parallel c$ (b).}
\end{figure*}

From our detailed dilatometric and thermodynamic data we construct the phase diagrams for the $c$- and $a$*-axes in Fig.~\ref{PD_All}. While the general features for $B\parallel c$ confirm previous results~\cite{Kurumaji2019, Hirschberger2020PRB, Zhang2020}, our data evidence two phases in zero-field which were previously unknown: (1) Our isothermal magnetization data between 19~K and 22~K (SI, Fig.~\ref{SI_Magnetometry}) clearly indicate that the IC-2 phase does not extend to zero field, but there is a separate pocket closed off by a phase boundary extending from the edge of the A(SkL) phase to \TNOne~=~22.3~K. We label this new phase IC-3, since incommensurate spin structures were previously reported for this temperature regime~\cite{Kurumaji2019}. (2) Furthermore, the phase boundary at $T^*$ splits the IC-1 phase into IC-1 and IC-1' (Fig.~\ref{alpha_0T}(b) inset). The yet unreported phase diagram for $B\parallel a$* in general shows a similar behavior, with the critical fields of the IC-1' and IC-4 phases at lowest temperatures assuming higher values than IC-1' and the A phase for the $c$-axis. The IC-4 phase appearing for $B\parallel a$* (see Fig.~\ref{PD_All}b) reminds of the A(SkL) phase for $B\parallel c$, however, it was shown previously by angle-dependent resistivity measurements in the $a$*-$c$-plane at 2~K that it does not connect to the A(SkL) phase.\cite{Hirschberger2019} The magnetic structure of this phase needs to be clarified by diffraction studies. One major difference between the phase diagrams is seen for $B\parallel a$*, where the IC-2 phase is not directly adjacent to the field-induced FM phase, but there is an additional phase (labelled B) in between (see Fig.~\ref{TE-and-M-of-T}(c) and (d)). The B phase is bordered by two jumps both in \cp\ as well as in ${\partial}\chi/{\partial}T$. 

From the anomalies in the thermal expansion, specific heat and magnetization at the phase boundaries we calculated the uniaxial pressure and field dependence of the critical temperatures and critical fields, as well as the entropy changes at the phase boundaries for $B\parallel c$-axis. Considering $T^*(B)$ marking a continuous phase transition, the associated jumps in specific heat ($\Delta$\cp ), magnetic susceptibility ($\Delta(\partial M/\partial B)$), and thermal expansion coefficient ($\Delta \alpha$) are connected with the magnetic field and pressure dependencies of $T^*$ by the Ehrenfest-type relations (see e.g.~Ref.~[\onlinecite{Barron-White1999, Klingeler2005}])

\begin{equation} \label{eq:Ehrenfest_1} 
\left.\left(\frac{{\partial}T^*}{{\partial}B_i}\right)\right|_p = -T^*\frac{\Delta (\left.\frac{\partial M_i}{\partial T})\right|_B}{\Delta c_{p,B}}
\end{equation}
and
\begin{equation} \label{eq:Ehrenfest_2} 
\left.\left(\frac{{\partial}T^*}{{\partial}p_i}\right)\right|_B = T^*V_{\mathrm{m}}\frac{\Delta\alpha_i}{\Delta c_{\mathrm{p}}}.
\end{equation}

Using the molar volume $V_{\mathrm{m}} = 7.06\cdot 10^{-5}$~m$^3$/mol (see Ref.~[\onlinecite{Frontzek2009}]) as well as the anomaly values presented above and in Tab.~\ref{tab:T-star} -- for their extraction from the experimental data see the supplement, Fig.~\ref{SI_T-star} -- we obtain a moderate uniaxial pressure dependence of ${\partial}T^*/{\partial}p_{\mathrm{c}} = -1.4(3)$~K/GPa. In a field of 0.2~T, the field dependence is very small and amounts to only ${\partial}T^*/{\partial}B_{\mathrm{c}} = -37(13)$~mK/T. At 0.3~T, ${\partial}T^*/{\partial}B_{\mathrm{c}}$ rises to $-0.28(18)$~K/T. These results demonstrate that the IC-1 phase is stabilized both under pressure and applied field at the expense of the IC-1' phase. Also, from these values we can calculate the expected jump in $\partial M/\partial B$ ($\Delta\chi$), at $T^{*}(B)$ via

\begin{equation}
    \left.\left(\frac{{\partial}T^*}{{\partial}B}\right)\right|_p = -\frac{\Delta\left(\frac{{\partial}M}{{\partial}B}\right)|_{p,T}}{\Delta\left(\frac{{\partial}M}{{\partial}T}\right)|_{p,B}}.
\end{equation} 

At 0.2~T this yields $\Delta\chi = -2\cdot 10^{-4} \mu_{\mathrm{B}}$/(f.u. T), which is well below the resolution limit of our experiment, explaining why our isothermal magnetization studies do not show anomalies at $T^*$ (see SI, Fig.~\ref{SI_T-star}). Note, that in an early report on \GPS~by Mallik et al.~the authors detected a jump in the effective local field $|$B$_{\mathrm{eff}}|$ at 15~K by M\"{o}ssbauer spectroscopy and attributed it to a lower ordering temperature of one of the two Gd sites in \GPS. This transition was not detected in any of the later reports on single crystalline samples, but our results presented in this work clearly confirm its presence. 

The phase boundaries between the A(SkL), IC-2, and DP phases are of discontinuous nature, exhibiting jumps $\Delta(dL_i/L_i)$ in the length changes and $\Delta M$ in the magnetization (i.e., $\Delta m$ in magnetic moment). Therefore, the Clapeyron equations~\cite{Barron-White1999}

\begin{align}\label{eq:Clapeyron}
    \left.\left(\frac{{\partial}T_c}{{\partial}p_i}\right)\right|_B &= V_{\mathrm{m}}\frac{\frac{{\Delta}L_i}{L_i}}{\Delta S} \\
    \left.\left(\frac{{\partial}T_c}{{\partial}B_i}\right)\right|_p &= -\frac{\Delta m_i}{\Delta S} = -\frac{\Delta (M_i\cdot V)}{\Delta S}\\
    \left.\left(\frac{{\partial}B_c}{{\partial}p_i}\right)\right|_T &= V_{\mathrm{m}}\frac{\frac{{\Delta}L_i}{L_i}}{\Delta m_i}
\end{align}

apply for the pressure and field dependence of the respective critical temperatures $T_{\mathrm{c}}$ and critical fields $B_{\mathrm{c}}$. Hence, the observed slopes ${\partial}T_c/{\partial}B$ (see Tab.~\ref{tab:IC-1-to-A}--\ref{tab:IC-2-to-DP}) and the jumps $\Delta M$ allow us to obtain the associated entropy changes $\Delta S$. 


For the transition from the incommensurate IC-1' phase to the skyrmion lattice A phase, a jump $\Delta M = 0.99(10)\mu_{\mathrm{B}}/$Gd accompanies the $c$-axis contraction of $\Delta dL_{\mathrm{c}}/L_{\mathrm{c}} = -10.8(1.1)\cdot 10^{-6}$ at 4.3~K. Applying the above-mentioned thermodynamic relations yields small entropy changes on the order of $\Delta S_{\mathrm{calc}} = 125(13)$~mJ/mol~K and a negative uniaxial pressure dependence of ${\partial}T_{\mathrm{c}}/{\partial}p_i = -6.1(9)$~K/GPa. At higher temperatures these values decrease down to $110(11)$~mJ/mol K and $-1.5(2)$~K/GPa at 16~K (see Tab.~\ref{tab:IC-1-to-A}). At the transition from the skyrmion lattice A phase to the IC-2 phase the $c$-axis also contracts, but these contractions are much smaller ($\Delta dL_{\mathrm{c}}/L_{\mathrm{c}} = -2.4(3)\cdot 10^{-6}$ at 4.3~K) while the jumps in magnetization again roughly correspond to one Bohr magneton per Gd ion ($\Delta M = 1.01(11)\mu_{\mathrm{B}}/$Gd at 4.3~K). Accordingly, this phase boundary shows much smaller pressure dependence, i.e., ${\partial}T_{\mathrm{c}}/{\partial}p_i = -0.47(7)$~K/GPa at 4.3~K (see Tab.~\ref{tab:A-to-IC-2_Field}). 
At the same time, the steeper slope of the phase boundary $B_{\mathrm{c}}(T)$ implies larger changes in entropy of $360(40)$~mJ/mol K at 4.3~K which increases to almost $600(60)$~mJ/mol K at 16~K.
We note that the analysis of anomalies from temperature instead of field sweeps confirms these values (Tab.~\ref{tab:A-to-IC-2_Temp}). 
Looking at higher fields, the slope of the phase boundary from IC-2 to the depinned phase (DP) is very small, changing from a small negative slope below 6~K to a small positive slope above. Considering the measured anomalies $\Delta M = 0.08(4)\mu_{\mathrm{B}}/$Gd at 4.3~K, this yields negligible associated entropy changes (Tab.~\ref{tab:IC-2-to-DP}). In contrast, there are pronounced lattice effects ($\Delta dL_{\mathrm{c}}/L_{\mathrm{c}} = -17(2)\cdot 10^{-6}$ at 1.77~K) yielding a very large pressure dependence for the phase boundary IC-2 $\rightarrow$ DP.

\begin{table*}[tb]
    \centering
    \begin{tabular}{c|cccccccc}
    \hline \hline
             & $T$ & $B_c$ & $\Delta(dL/L)$ & $\Delta m$  & ${\partial}B_{\mathrm{c}}/{\partial}T$  & $\Delta S$ & ${\partial}T_{\mathrm{c}}/{\partial}p_{\mathrm{i}}$  & ${\partial}B_{\mathrm{c}}/{\partial}p_i$ \\ 

             Transition & (K) & (T) & (10$^{-6}$) & ($\mu_{\mathrm{B}}$/Gd) & (T/K) & (mJ/mol K) & (K/GPa) & (mT/GPa) \\ \hline

        IC-1 $\rightarrow$ A(SkL) & 4.3$\pm$0.1 & 0.49$\pm$0.02 & --10.8$\pm$1.1 & 1.0$\pm$0.1 & --0.01 & 125$\pm$13 & --6.1$\pm$0.9 & --50$\pm$50 \\ 
        A(SkL) $\rightarrow$ IC-2 & 4.3$\pm$0.1 & 1.04$\pm$0.03 & --2.4$\pm$0.3 & 1.0$\pm$0.1 & --0.03 & 360$\pm$40 & --0.47$\pm$0.07 & --15$\pm$2 \\ 
        \hline
    \end{tabular}
    \caption{\label{tab:Results}Relevant quantities and anomaly sizes at the phase boundaries of the skyrmion lattice phase, at $T=4.3$~K, and in magnetic fields $B\parallel c$ which have been either directly extracted from the experimental data or were obtained by using thermodynamic relations as given in the text.} 
\end{table*}

The results of the thermodynamic analyses are shown in table I as well in tables~\ref{tab:IC-1-to-A} to \ref{tab:IC-2-to-DP} in the supplement. In particular, our analysis evidences pronounced negative uniaxial pressure dependencies for all three phase transitions between IC-1/IC-1', A(SkL), IC-2 and DP at low temperatures. This implies that the IC-1'/IC-1, A(SkL) and IC-2 phases are all destabilized by pressure along the $c$-axis with respect to the higher temperature phases, i.e., the field-induced FM phase -- and the paramagnetic phase at low fields -- is stabilized.

In particular, our data for $B\parallel c$ provide further information on the skyrmion phase. Both the onset of the SkL phase from incommensurate magnetic order IC-1/IC-1' and its transition into the incommensurate IC-2 phase are discontinuous in nature. In both cases, transitions are associated with the increase of magnetization by about 1~$\mu_{\mathrm{B}}/$Gd. Rather flat phase boundaries in the magnetic phase diagram are indicative of comparably small entropy changes. Our quantitative analysis evidences that the evolution of the SkL phase, depending on the temperature, yields an entropy gain of $\Delta S\approx 100-150$~mJ/(mol K) while the entropy jumps at the transition out of the skyrmion phase by 300-600~mJ/(mol K). These values are by far larger than for the chiral magnet \ce{MnSi} where latent heat at the phase boundaries only amount to a few m\jmk .~\cite{Bauer2013} Uniaxial pressure along the $c$-axis significantly enhances the SkL phase as seen by the uniaxial pressure dependencies of the transition temperatures. Specifically, at 4~K there is a rapid decrease of the IC-1/IC-1' $\rightarrow$ A(SkL) transition temperature $\partial T_{\mathrm{in}}/\partial p_c\approx -6$~K/GPa, leading to an expansion of the A(SkL) phase towards lower temperatures under pressure. At the same time the temperature of the A(SkL) $\rightarrow$ IC-2 transition, i.e. exiting the SkL phase towards higher temperatures, changes by only $\partial T_{\mathrm{out}}/\partial p_c\approx -0.5$~K/GPa (see tables I and II). Enhancement of skyrmion lattice phases under pressure is also observed in other materials. In the insulating skyrmion system Cu$_2$OSeO$_3$, Levati\'c et al.~report a dramatic enhancement of the skymion pocket under pressure by about 8~K at 0.6~GPa.~\cite{Levatic2016} While in \GPS\ the SkL phase appears at lower temperatures, our results ($\partial T_{\mathrm{in}}/\partial p_c-\partial T_{\mathrm{out}}/\partial p_c$) imply about half of this effect. We also note similar findings to the ones reported at hand in the chiral magnet MnSi~\cite{Chacon2015,Nii2015} where uniaxial pressure along [001] yields a rapid decrease of the onset temperature of the skyrmion phase while the high temperature phase boundary shows a much smaller pressure dependence.~\cite{FN3}
Recent theoretical studies by Hayami et al.~investigated the influence of single-ion anisotropy on the formation and stability of the skyrmion lattice phase. They show that easy-axis anisotropies stabilize magnetic-field-induced skyrmion crystals in frustrated magnets\cite{Hayami2016} and easy-axis (easy-plane) anisotropy substantially increases (decreases) the stable field-range for a Skyrmion lattice. 
These findings suggest, that the pressure dependencies stabilizing the skyrmion lattice phase in \GPS\ may originate from small distortions in the local environment of \ce{Gd} leading to an increase of the weak magnetic anisotropy of the Gd moments.

The transition from the depinned phase to the field-induced ferromagnetic phase is of a continuous type. It exhibits a jump in the magnetostriction coefficient $\Delta \lambda = -4.7(5)\cdot 10^{-5}/T$ (at 1.77~K) and in the derivative of the magnetization $\Delta \partial M/\partial B = -0.159(16) \mu_{\mathrm{B}}$/(T Gd) (T~=~1.9~K). Using an Ehrenfest relation, the uniaxial pressure dependence of the critical field can be expressed as $dB_{\mathrm{c}}/dp_i = \Delta \lambda / \Delta(dM/dB)$, which yields $dB_{\mathrm{c}}/dp_c = 1.9(5)$~T/GPa at 1.77~K, i.e. the depinned phase is stabilized under uniaxial pressure $p_c$. Similarly, using the anomaly values listed in Tab.~\ref{tab:IC-2-to-fiFM}, we find uniaxial pressure dependencies of the critical field between $-0.65(16)$~T/GPa (10~K) and $-0.50(8)$~T/GPa (14~K) at the continuous transition IC-2 to fiFM.



While \GPS\ shows only moderate frustration, magnetic entropy and length changes are observed up to about 2.7~\TNOne\ (60~K), thereby implying the evolution of short range magnetic order in this temperature regime.
Effects of fluctuations above \TN\ in \GPS\ were observed before in resistivity measurements. Measurements on polycrystalline samples show a well-defined minimum around 45~K~\cite{Mallik1998EPL} which was also confirmed in single crystals~\cite{Saha1999}. A theoretical explanation of this behavior based on the RKKY-interaction in combination with frustration was given by Wang et al.\cite{Wang2016}.
Gr\"{u}neisen scaling suggests that these precursor fluctuations are of the IC-1/IC-3 type. Both ordering phenomena are driven by the same dominating energy scale which differs from the one driving IC-1'. 
As expected for a Gd$^{3+}$-system, magnetoelastic coupling is moderate. It is, hence, somehow surprising that magnetostriction is large at high temperatures and displays pronounced effects up to 200~K. In addition, despite linear field dependence of the magnetization, magnetostriction does not follow a $B^2$-law below 200~K (see Fig.~\ref{SI_dchidp}) as would be expected from the relation $dL_i/L_i = -1/2 V\partial \chi_i/ \partial p_i B^2$ in the paramagnetic regime~[\onlinecite{Johannsen2015}]. Tentatively, magnetostriction above 100~K implies negative uniaxial pressure dependence, $\partial\chi/\partial p_{c}<0$, of the magnetic susceptibility while $\partial\chi/\partial p_{\mathrm{a^*}}>0$. 
This observation suggests that antiferromagnetic exchange interactions are strengthened by uniaxial pressure along the $c$-axis and weakened upon application of $p\parallel a$*. Notably, however, the long-range magnetic ordering temperatures do not follow this trend as $\partial T_{\mathrm{N}}/\partial p_c<0$ which further highlights the complex nature of magnetism in \GPS .

\section{Conclusions}
For the first time, high-resolution dilatometry was used to study the interplay between magnetism and the lattice of single crystalline \GPS. Strong magnetoelastic coupling and field effects up to high temperatures are found. Pronounced anomalies in the thermal expansion, magnetostriction and magnetization allow us to obtain the magnetic phase diagram. This yields in particular several novel phases for $B\parallel c$ while the B vs. T phase diagram for $B\parallel a$* has not yet been reported at all in the literature. Gr\"{u}neisen analysis shows the onset of magnetic contributions well above \TNOne , and the pressure dependencies of ordering phenomena are obtained. In particular, we find that uniaxial pressure strongly enhances the skyrmion lattice phase.

\section*{Acknowledgements}
We thank I. Mazilu and Y. Xu for support in the crystal growth. We acknowledge financial support by BMBF via the project SpinFun (13XP5088) and by Deutsche Forschungsgemeinschaft (DFG) under Germany’s Excellence Strategy EXC2181/1-390900948 (the Heidelberg STRUCTURES Excellence Cluster), through project KL 1824/13-1, and within the SFB 463.

\bibliographystyle{apsrev}
\bibliography{Gd2PdSi3_Bibliography}

\begin{thebibliography}{53}
\bibitem[{FN3()}]{FN3}
\emph{\bibinfo{title}{\rm{For MnSi, this also holds for pressure along [111]
  and [110]. From our data, only uniaxial pressure effects $p\parallel c$ are
  accessible so that a comparison for the other directions cannot be made.}}}
\end{thebibliography}

\clearpage
\newpage

\beginsupplement

\onecolumngrid

\section*{\Large{Supplementary Material:\\ Magnetoelastic Coupling and Phases in the Skyrmion Lattice Magnet \ce{Gd2PdSi3} Discovered by High-resolution Dilatometry}}

\begin{figure*}[htbp]
	\center{\includegraphics [width=0.9\columnwidth,clip]{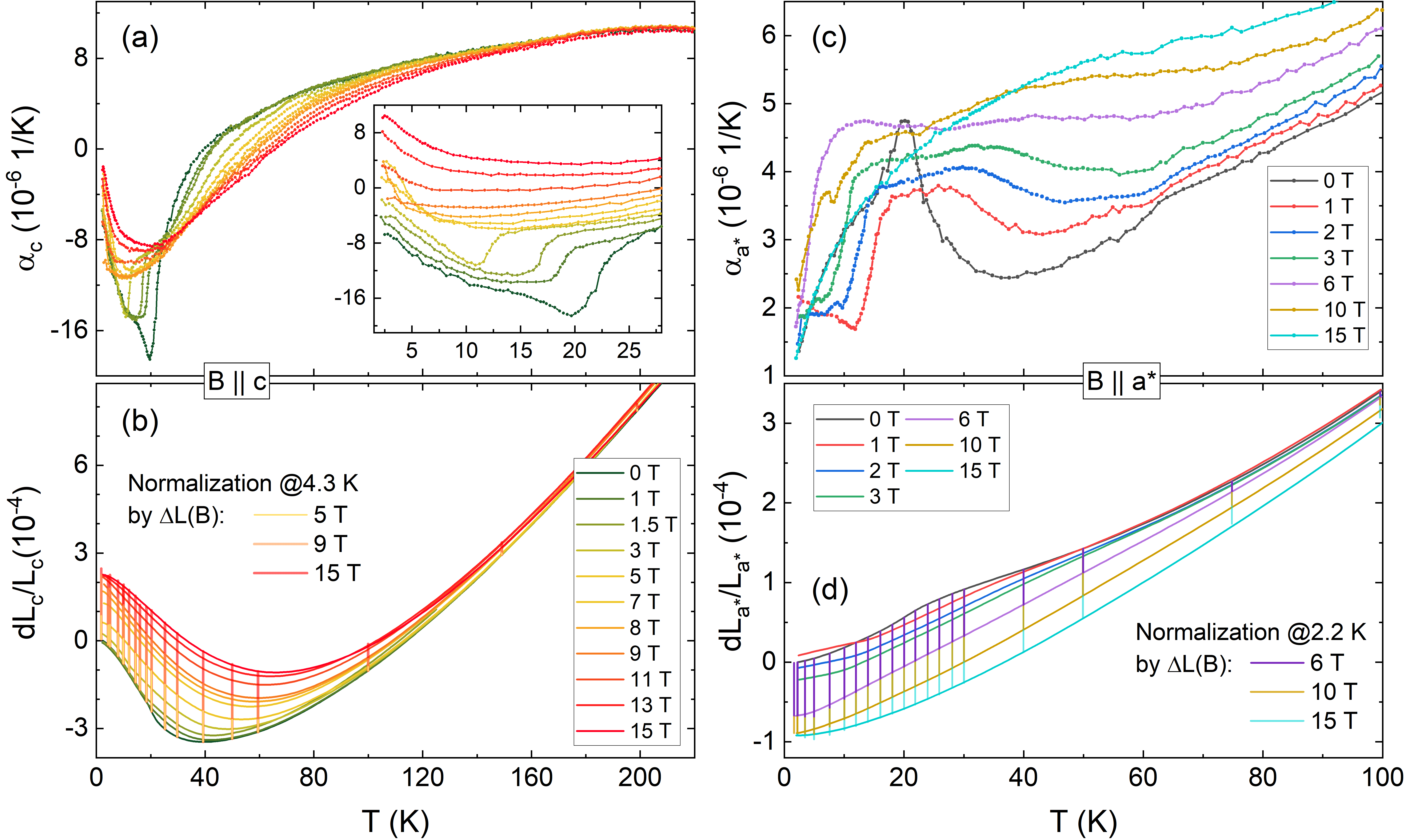}}
	\caption[] {\label{SI_TE} (a-b) Thermal expansion measurements up to 220~K for $B\parallel c$ in magnetic fields up to 15~T. (a) Thermal expansion coefficient \ac. The inset shows the magnified low temperature region with an offset of $1.2\cdot 10^{-6}$~1/K. (b) Relative length changes $dL_{\mathrm{c}}(T)/L_{\mathrm{c}}$. Vertical lines mark the changes $\Delta L(B)/L$ derived from magnetostriction measurements. The data at different fields are normalized by these changes at 4.3~K. (c-d) Thermal expansion measurements up to 100~K for $B\parallel a$* in magnetic fields up to 15~T. (c) Thermal expansion coefficient \ain\ offset by $2\cdot 10^{-7}$~1/K. (d) Relative length changes $dL_{\mathrm{a^*}}(T)/L_{\mathrm{a^*}}$. Vertical lines again mark the changes $\Delta L(B)/L$ taken from magnetostriction measurements. The data at different fields are offset and normalized by these changes at 2.2~K w.r.t. the zero-field data.}
\end{figure*}

\begin{figure*}[ht]
	\center{\includegraphics [width=0.8\columnwidth,clip]{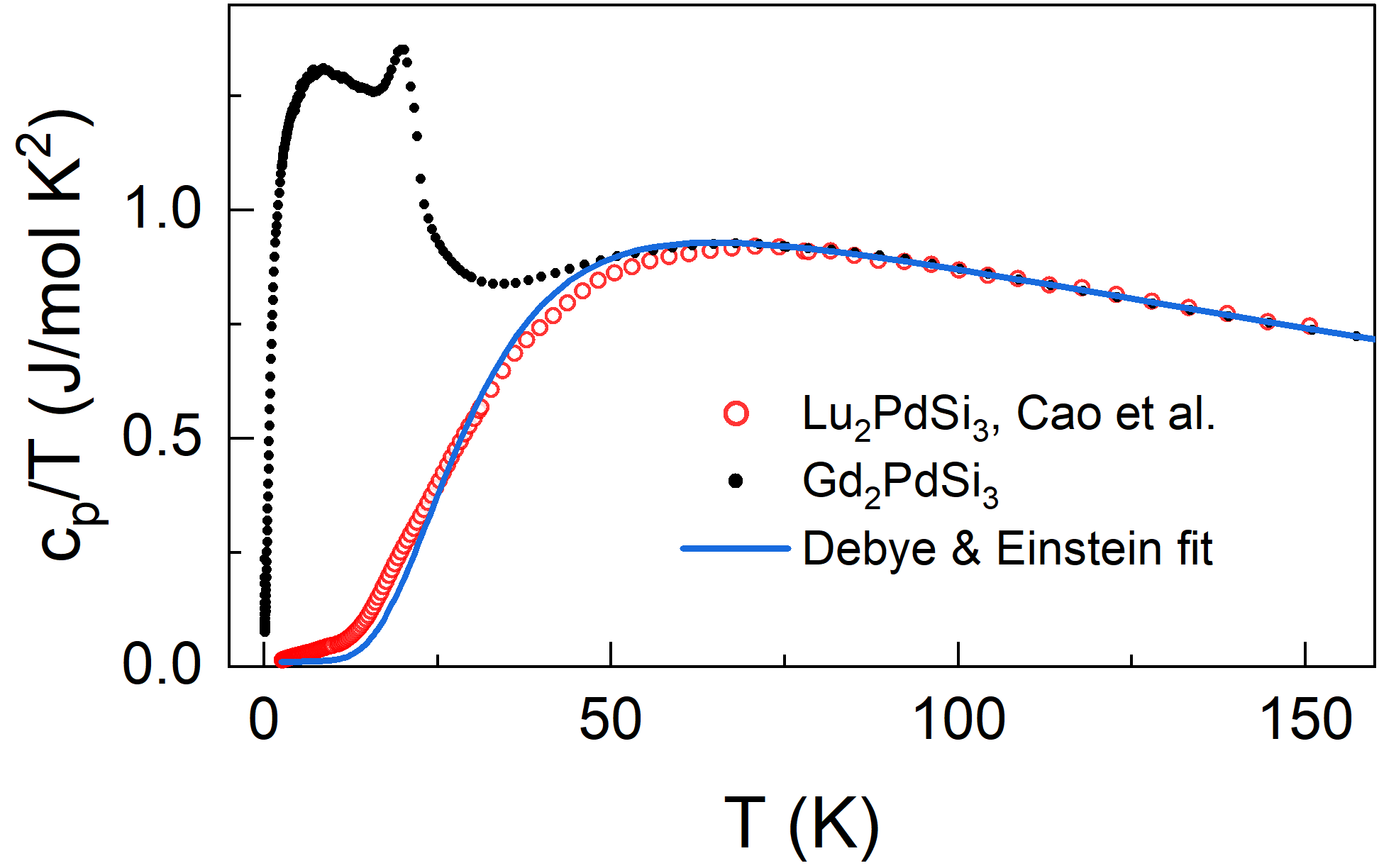}}
	\caption[] {\label{SI_cpt} Comparison of the scaled \LPS\ specific heat \cpt\ measured by Cao et al.\cite{Cao2013} (red empty circles) with the Debye and Einstein fit (blue line). Note that the \LPS\ data was interpolated and therefore contains more data points than the published data.}
\end{figure*}

\begin{figure*}[ht]
	\center{\includegraphics [width=0.9\columnwidth,clip]{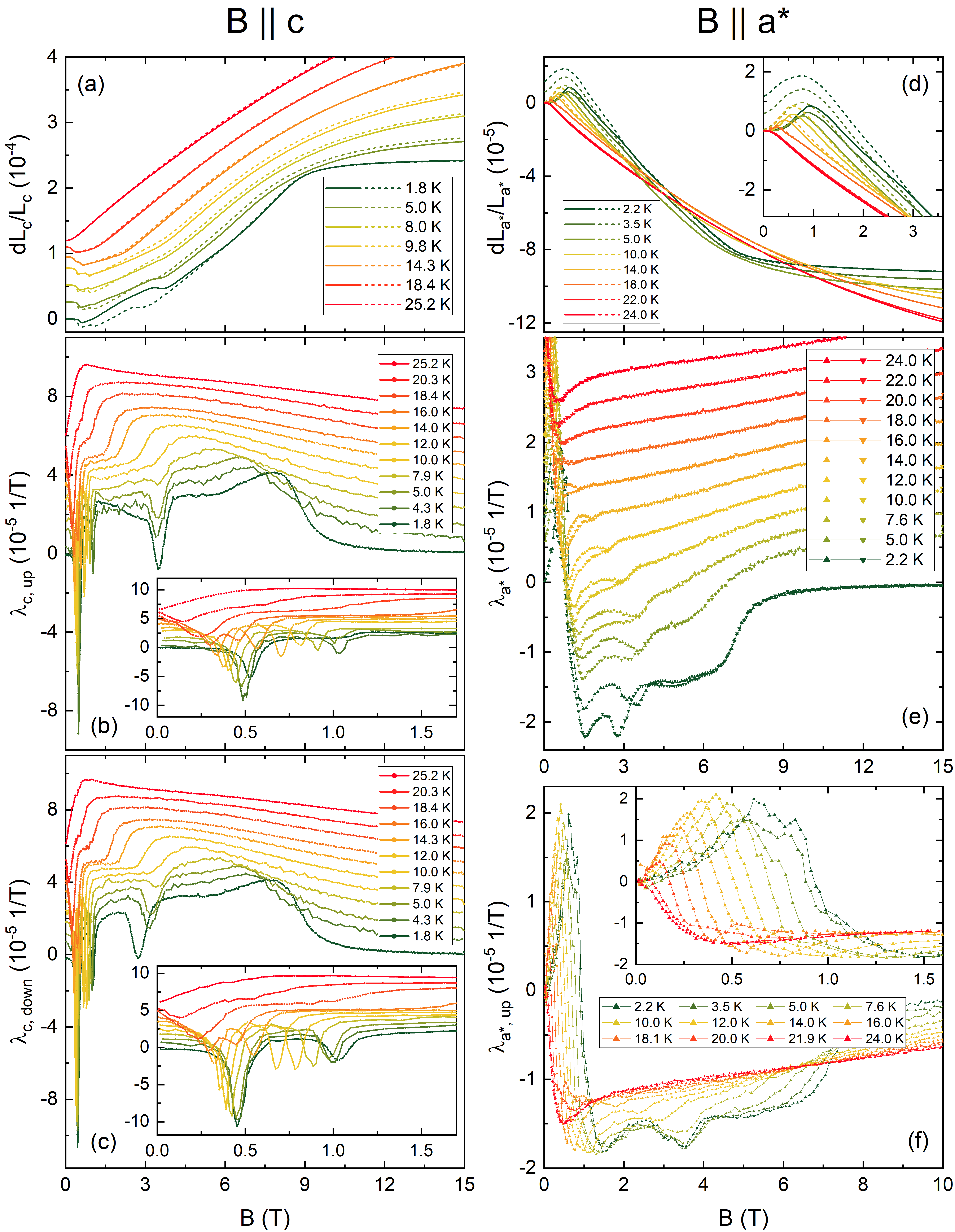}}
	\caption[] {\label{SI_MS_Low} Magnetostriction measurements at temperatures from 1.7~K to 25~K for $B\parallel c$ (a-c) and $B\parallel a$* (d-f): The relative length change $dL(B)/L$ (a, d) and magnetostriction coefficient $\lambda_i$ (b, c, e, f) are shown. Solid lines in (a, d) represent up-sweeps, dashed lines down-sweeps. Triangles pointing upwards in (d) mark up-sweeps, triangles pointing downwards mark down-sweeps. Data in (a-c, insets also) and (e) are shifted vertically for better visibility by: (a)  different amounts, (b, c) $6\cdot 10^{-6}$~1/T, (e) $3.7\cdot 10^{-6}$~1/T (3.5~K data is omitted). Insets show magnifications of the low field regions.}
\end{figure*}

\begin{figure*}[ht]
	\center{\includegraphics [width=0.9\columnwidth,clip]{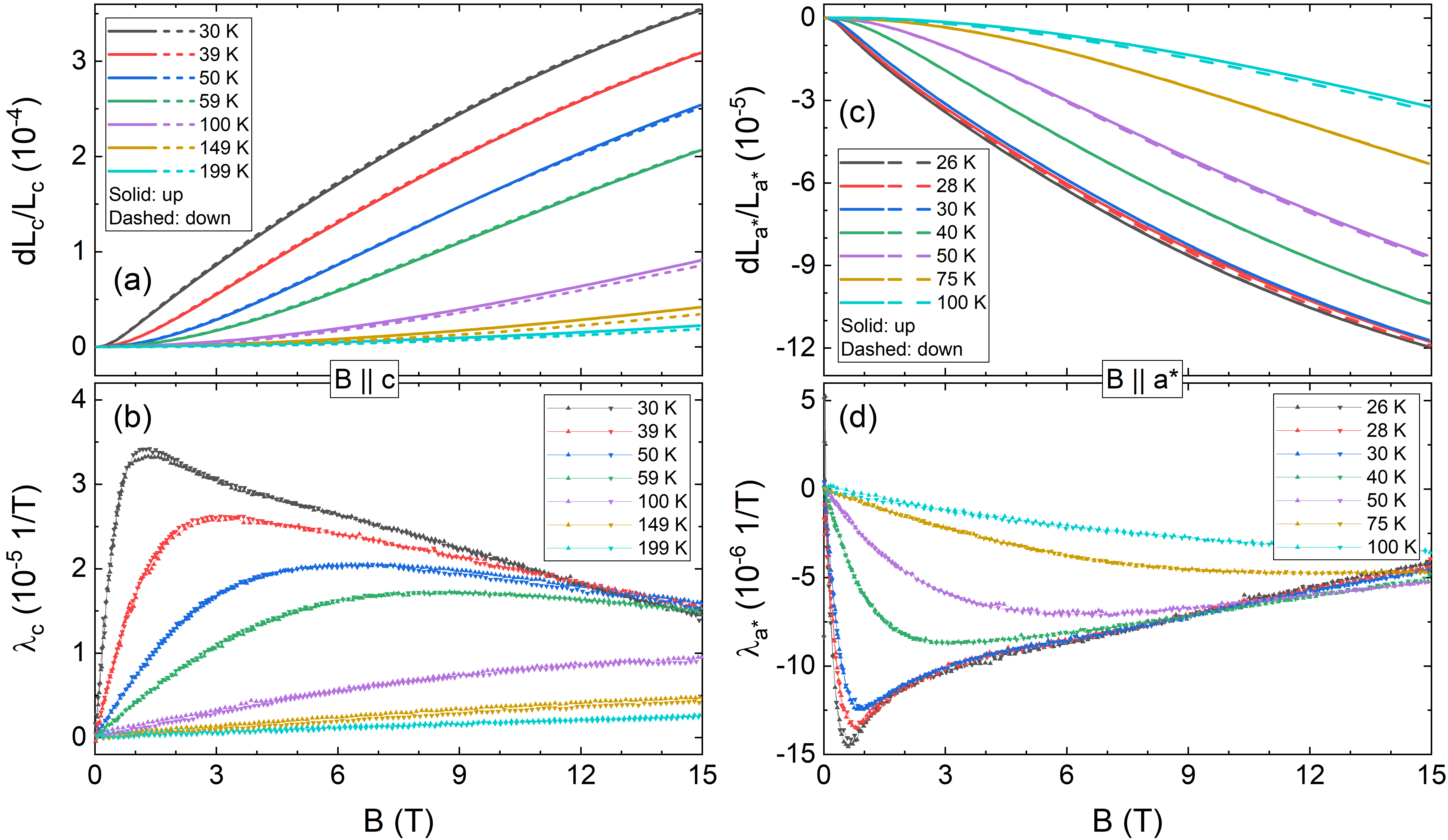}}
	\caption[] {\label{SI_MS_High} Magnetostriction measurements at temperatures above \TNOne~for $B\parallel c$ (a-b) and $B\parallel a$* (c-d): The relative length change $dL(B)/L$ (a, c) and magnetostriction coefficient $\lambda_i$ (b, d) are shown. Solid lines in (a, c) represent up-sweeps, dashed lines down-sweeps. Triangles pointing upwards in (b, d) are up-sweeps, triangles pointing downwards mark down-sweeps.}
\end{figure*}

\begin{figure*}[htbp]
	\center{\includegraphics [width=0.8\columnwidth,clip]{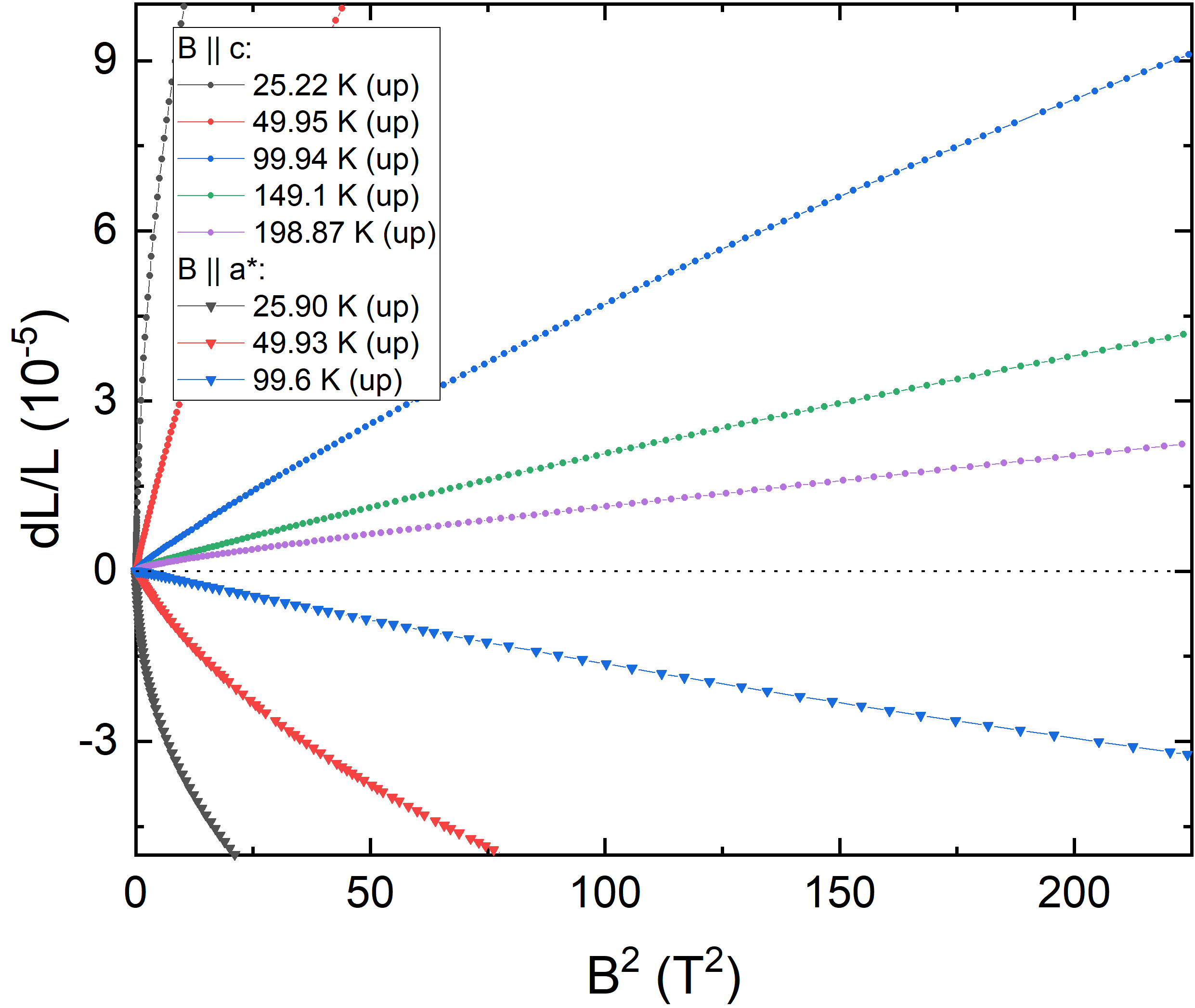}}
	\caption[] {\label{SI_dchidp} Relative length changes $dL/L$ plotted vs. $B^2$. The uniaxial pressure dependence of the susceptibility is related to the magnetostriction by the Maxwell relation ${\partial}(dL/L)/{\partial}B = -{\partial}M/{\partial}p$. For a paramagnetic material with $M = \chi B$ we thus have $dL/L = -\frac{1}{2} {\partial}\chi/{\partial}p B^2$, i.e. ${\partial}\chi/{\partial}p$ is proportional to the slope in the above plot.}
\end{figure*}

\begin{figure*}[ht]
	\center{\includegraphics [width=0.8\columnwidth,clip]{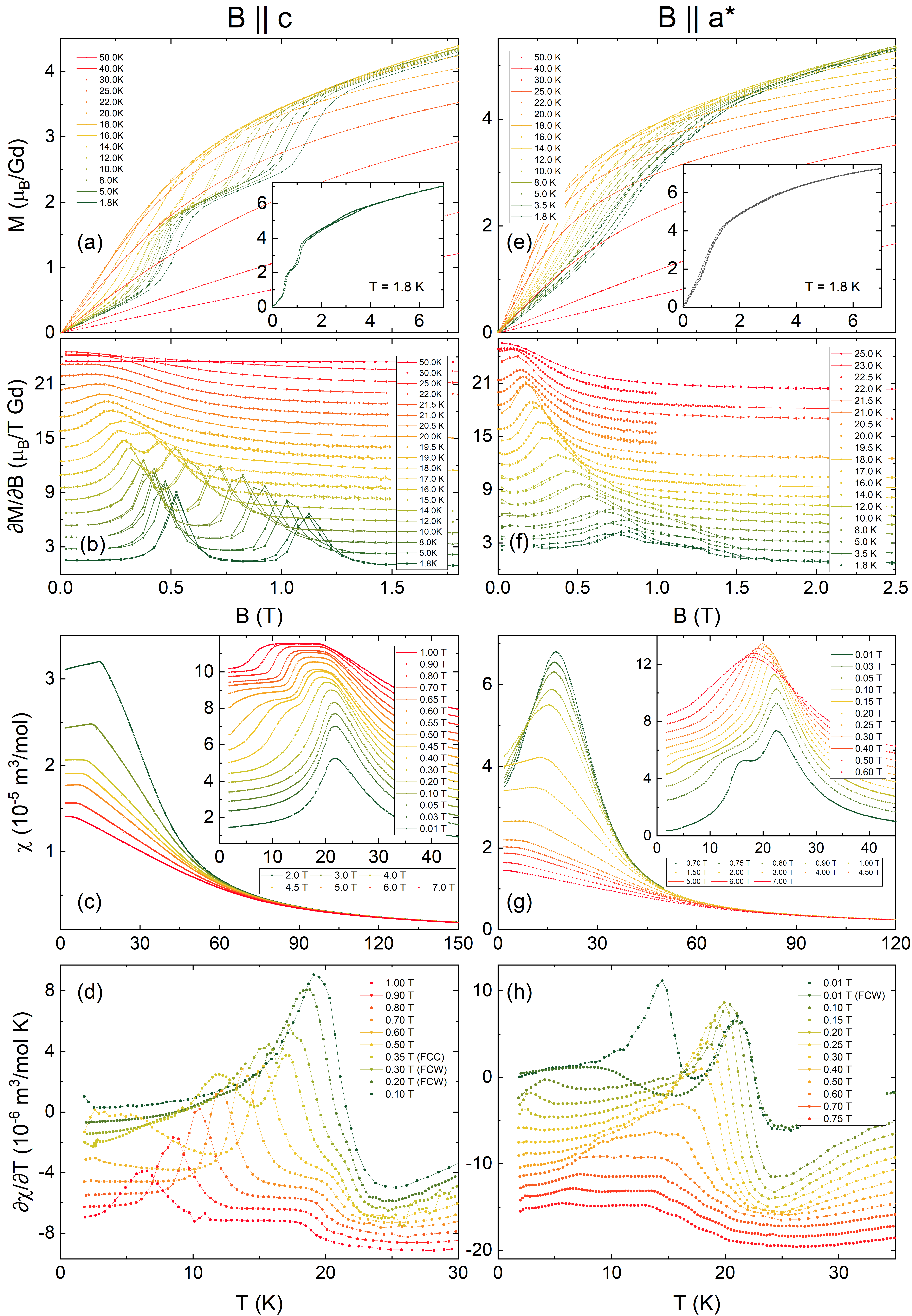}}
	\caption[] {\label{SI_Magnetometry} Measurements of the isothermal magnetization as well as the static magnetic susceptibility $\chi(T)$ and their derivatives for $B\parallel c$ (a-d) and $B\parallel a$* (e-h). $\chi(T)$ measurements were performed in a zero-field cooled manner unless stated otherwise (FCW: field-cooled warming, FCC: field-cooled cooling). Data are offset for better visibility by (b) 1.2~$\mu_{\mathrm{B}}$/(T Gd), (c, inset) $4.5\cdot 10^{-6}$~m$^3$/mol, (d) $-8\cdot 10^{-7}$~m$^3$/(mol K), (f) 1.1~$\mu_{\mathrm{B}}$/(T Gd), (g, inset) $5\cdot 10^{-6}$~m$^3$/mol, and (h) $-1.4\cdot 10^{-6}$~m$^3$/(mol K) starting with the data for $B$~=~0.10~T.}
\end{figure*}

\begin{figure*}[ht]
	\center{\includegraphics [width=0.8\columnwidth,clip]{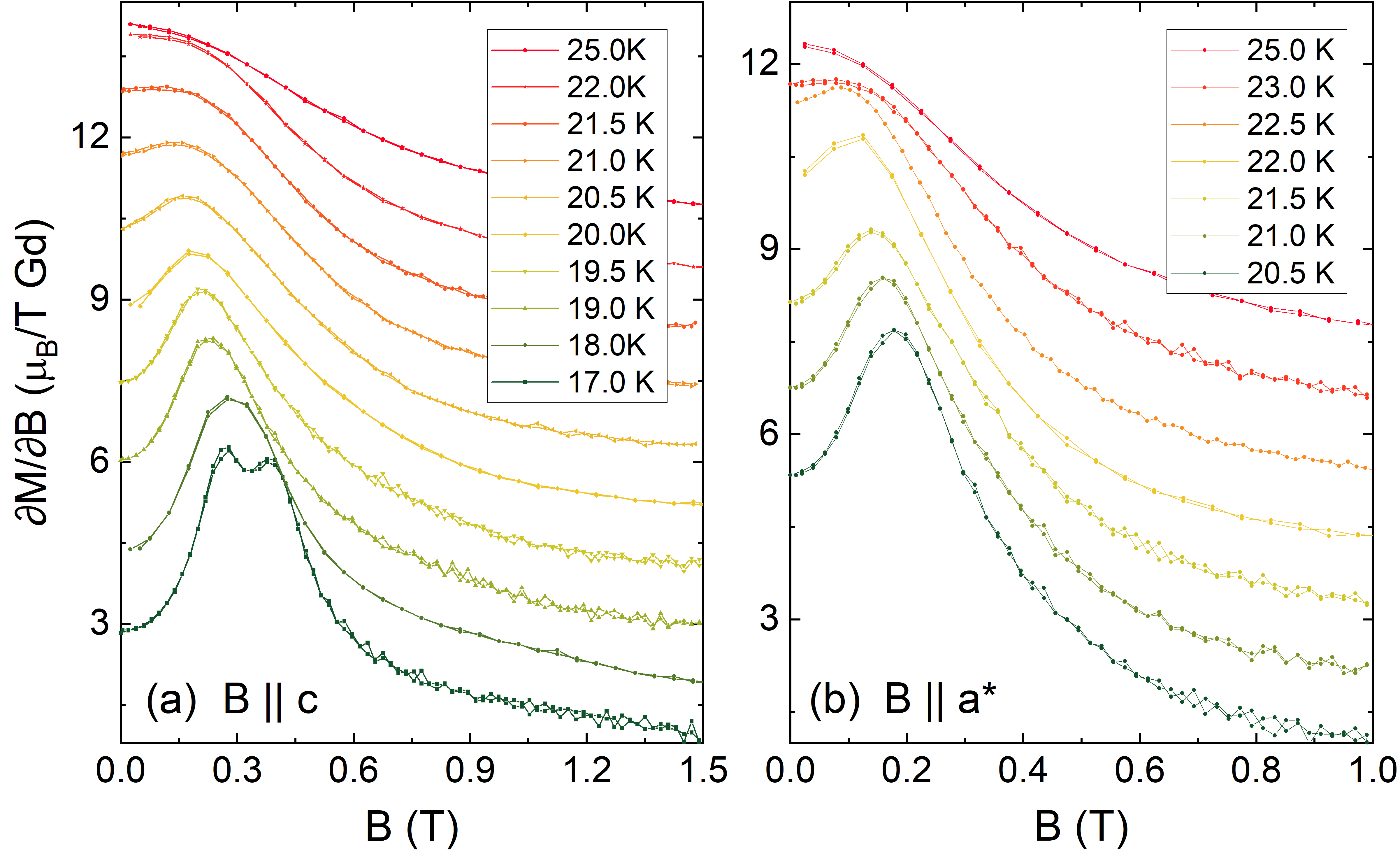}}
	\caption[] {\label{SI_dMdB} Close-up of the measurements of the magnetic susceptibility ${\partial}M/{\partial}B$ for $B\parallel c$ (a) and $B\parallel a$* (b) in the temperature regime of the IC-3 $\rightarrow$ IC-2 transition.}
\end{figure*}

\begin{figure*}[ht]
	\center{\includegraphics [width=0.6\columnwidth,clip]{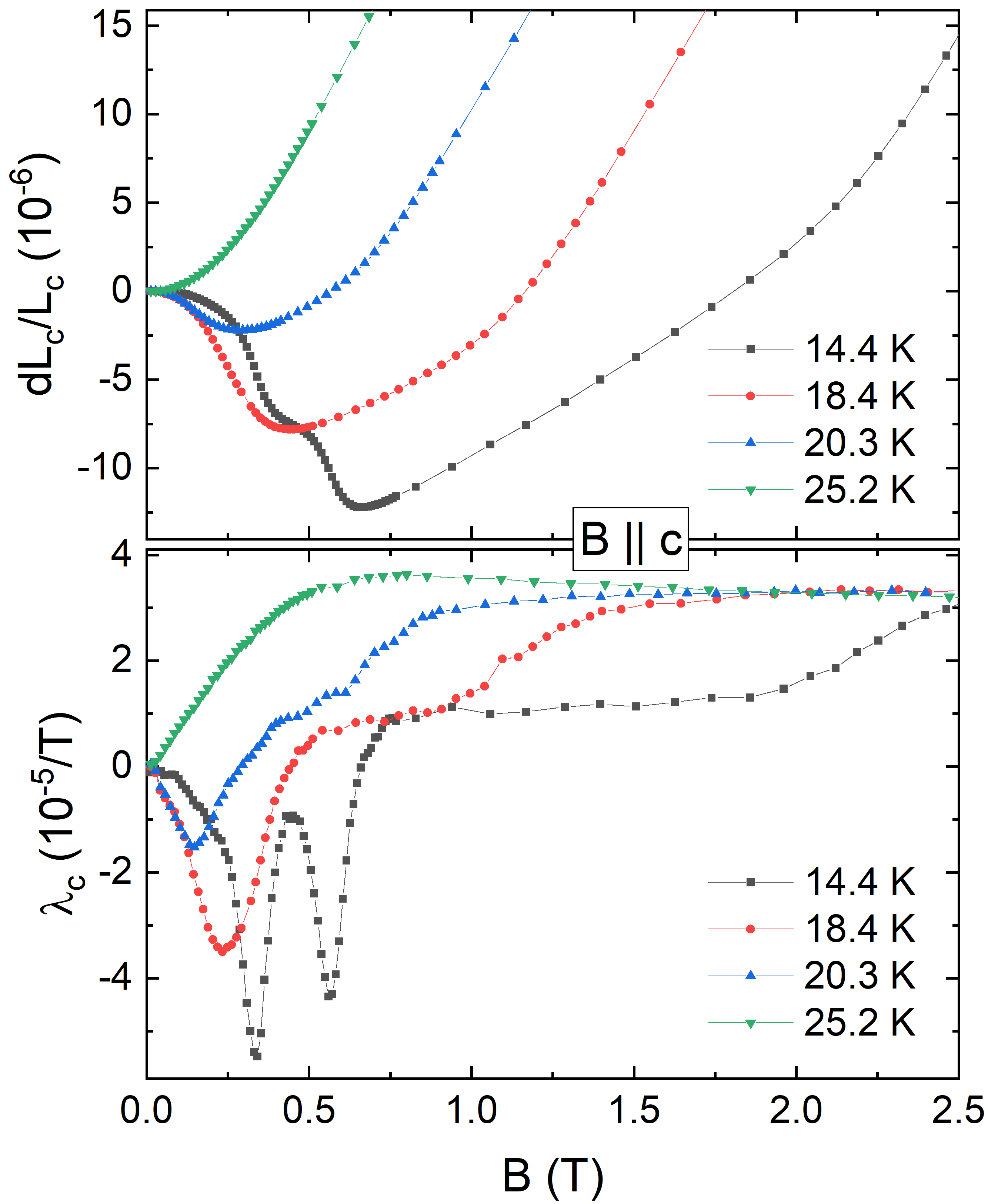}}
	\caption[] {\label{SI_Lambda_c} (a) Relative length changes $dL_{\mathrm{c}}/L_{\mathrm{c}}$ and (b) magnetostriction coefficient \lc\ in the low-field regime at temperatures below and above \TNOne\ and \TNTwo\ for $B\parallel c$. Only up-sweeps are shown.}
\end{figure*}

\begin{figure*}[ht]
	\center{\includegraphics [width=1\columnwidth,clip]{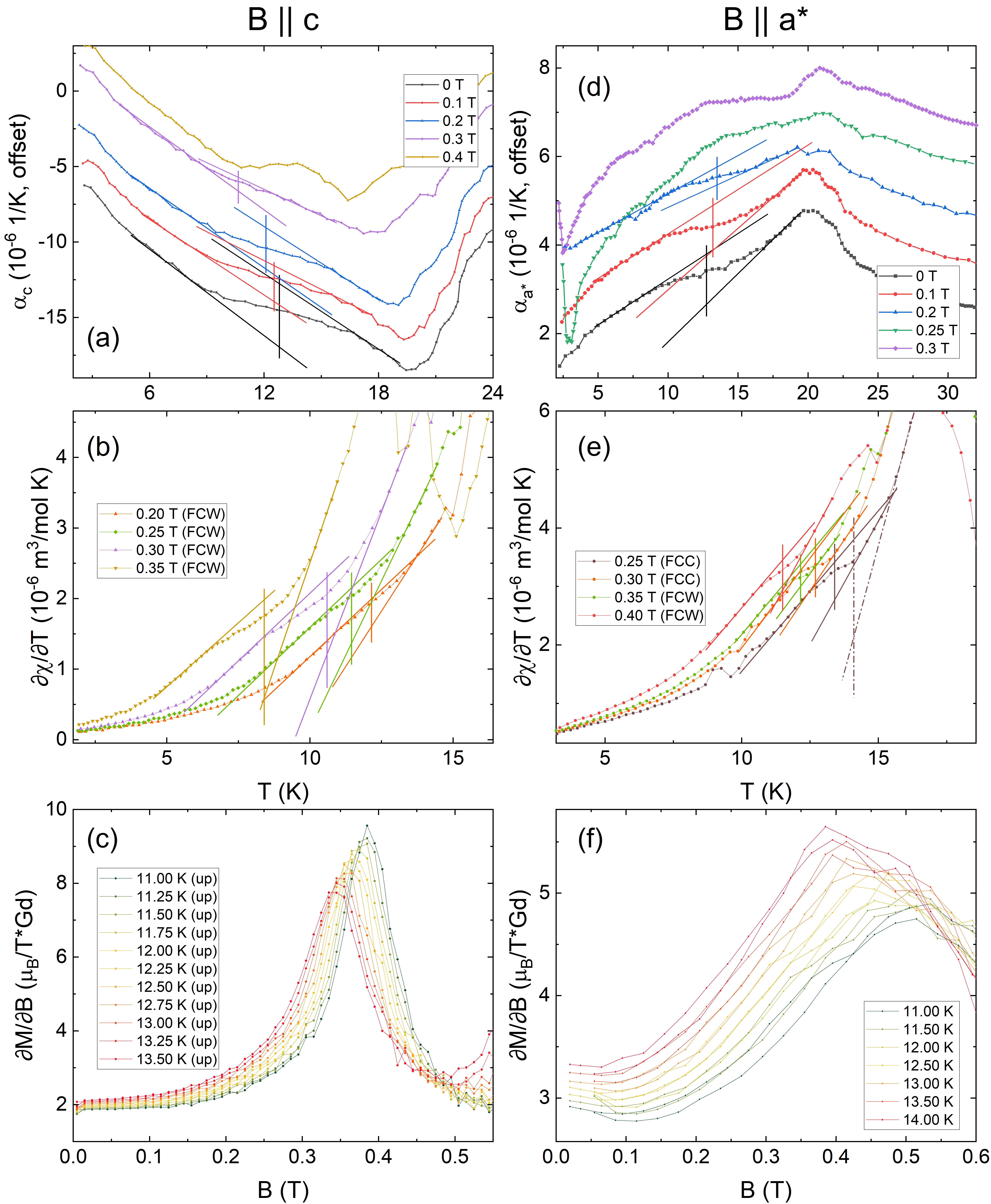}}
	\caption[] {\label{SI_T-star} Extraction of $T^*$ from thermal expansion (a, d), ${\partial}\chi/{\partial}T$ (b, e) and ${\partial}M/{\partial}B$ for $B\parallel c$ (a-c) and $B\parallel a$* (d-f). Vertical lines mark the position of $T^*$ for the different measurements. Data in (a) and (d) are offset by $2\cdot 10^{-6}$~1/K and $1\cdot 10^{-6}$~1/K, respectively.}
\end{figure*}

\begin{table*}[ht]
    \centering
    \begin{tabular}{cccccccc}
    \hline \hline
    \multicolumn{8}{c}{IC-1/IC-1' $\rightarrow$ A(SkL) -- $c$-axis: Relevant quantities, entropy changes and uniaxial pressure dependencies} \\
        T (K) & B$_{\mathrm{c}}$ (T) & $\Delta(dL/L)$ (10$^{-6}$) & $\Delta m$ ($\mu_{\mathrm{B}}$/Gd) & ${\partial}B_{\mathrm{c}}/{\partial}T$ (T/K) & $\Delta S_{\mathrm{calc}}$ (mJ/mol K) & ${\partial}T_{\mathrm{c}}/{\partial}p_{\mathrm{i}}$ (K/GPa) & ${\partial}B_{\mathrm{c}}/{\partial}p_i$ (T/GPa) \\ \hline
        1.77 & 0.52$\pm$0.02 & --7.3$\pm$0.8 & 1.00$\pm$0.10 & --0.01135 & 126$\pm$13 & --4.1$\pm$0.6 & --0.046$\pm$0.007\\ 
        4.3 & 0.49$\pm$0.02 & --10.8$\pm$1.1 & 0.99$\pm$0.10 & --0.01133 & 125$\pm$13 & --6.1$\pm$0.9 & --0.069$\pm$0.010\\ 
        5.01 & 0.48$\pm$0.02 & --8.9$\pm$0.9 & 0.99$\pm$0.10 & --0.01151 & 127$\pm$13 & --4.9$\pm$0.7 & --0.056$\pm$0.008 \\ 
        7.94 & 0.45$\pm$0.02 & --8.4$\pm$0.9 & 0.95$\pm$0.10 & --0.01318 & 140$\pm$14 & --4.2$\pm$0.6 & --0.056$\pm$0.008 \\ 
        9.83 & 0.42$\pm$0.02 & --8.7$\pm$0.9 & 0.81$\pm$0.08 & --0.01499 & 135$\pm$14 & --4.6$\pm$0.7 & -- 0.068$\pm$0.010 \\ 
        9.99 & 0.42$\pm$0.02 & --7.2$\pm$0.8 & 0.81$\pm$0.08 & --0.01516 & 137$\pm$14 & --3.7$\pm$0.6 & --0.056$\pm$0.008\\ 
        12.02 & 0.38$\pm$0.02 & --7.0$\pm$0.7 & 0.75$\pm$0.08 & --0.01783 & 149$\pm$15 & --3.3$\pm$0.5 & --0.059$\pm$0.009 \\ 
        14.35 & 0.34$\pm$0.02 & --4.5$\pm$0.5 & 0.52$\pm$0.05 & --0.02260 & 131$\pm$14 & --2.4$\pm$0.4 & --0.055$\pm$0.008 \\
        16 & 0.30$\pm$0.02 & --2.4$\pm$0.3 & 0.40$\pm$0.04  & --0.02505 & 110$\pm$11 & --1.5$\pm$0.2 & --0.038$\pm$0.006\\ \hline
        \multicolumn{8}{c}{IC-3 $\rightarrow$ IC-2 -- $c$-axis: Relevant quantities, entropy changes and uniaxial pressure dependencies} \\
        18.39 & 0.30$\pm$0.04 & --9$\pm$2 & 0.24$\pm$0.03  & --0.1 & 270$\pm$40 & --2.3$\pm$0.6 & --0.23$\pm$0.06\\
        20.27 & 0.23$\pm$0.04 & --3.8$\pm$0.4 & 0.21$\pm$0.05 & -0.04 & 94$\pm$30 & --2.9$\pm$0.8 & --0.12$\pm$0.03\\ \hline
    \end{tabular}
    \caption{\label{tab:IC-1-to-A}Relevant quantities and jumps, calculated changes of entropy, and calculated pressure dependencies for the discontinuous transition in magnetic field from IC-1/IC-1' $\rightarrow$ A(SkL) and IC-3 $\rightarrow$ IC-2 for the $c$-axis according to Eq.~\eqref{eq:Clapeyron}. The column of ${\partial}B_{\mathrm{c}}/{\partial}T$ was calculated by taking the derivative of a polynomial fit to the values $B_{\mathrm{c}}(T)$ for IC-1/IC-1' $\rightarrow$ A(SkL) and a linear fit (between between 18.39~K and 19~K as well as between 19~K and 21~K) for IC-3 $\rightarrow$ IC-2.}
\end{table*}

\begin{table*}
    \centering
    \begin{tabular}{cccccccc}
    \hline \hline
       \multicolumn{8}{c}{A(SkL) $\rightarrow$ IC-2 -- $c$-axis: Relevant quantities, entropy changes and uniaxial pressure dependencies} \\
        T (K) & B$_{\mathrm{c}}$ (T) & $\Delta(dL/L)$ (10$^{-6}$) & $\Delta m$ ($\mu_{\mathrm{B}}$/Gd) & ${\partial}B_{\mathrm{c}}/{\partial}T$ (T/K) & $\Delta S_{\mathrm{calc}}$ (mJ/mol K) & $dT_{\mathrm{c}}/dp_{\mathrm{i}}$ (K/GPa) & ${\partial}B_{\mathrm{c}}/{\partial}p_i$ (T/GPa) \\ \hline
        1.77 & 1.11$\pm$0.03 & 0  & 0.95$\pm$0.10 & --0.028 & 300$\pm$30 & 0 & 0 \\ 
        4.3 & 1.04$\pm$0.03 & --2.4$\pm$0.3 & 1.01$\pm$0.11 & --0.031 & 360$\pm$40 & --0.47$\pm$0.07 & --0.015$\pm$0.002 \\ 
        5.01 & 1.01$\pm$0.02 & --1.4$\pm$0.2 & 1.00$\pm$0.10 & --0.033 & 370$\pm$40 & --0.27$\pm$0.04 & --0.009$\pm$0.002\\ 
        7.94 & 0.91$\pm$0.02 & --2.5$\pm$0.3 & 0.99$\pm$0.10 & --0.041 & 450$\pm$50 & --0.40$\pm$0.06 & --0.016$\pm$0.003\\
        9.83 & 0.82$\pm$0.02 & --5.1$\pm$0.5 & 0.98$\pm$0.10 & --0.048 & 520$\pm$60 & --0.69$\pm$0.10 & --0.033$\pm$0.005\\
        9.99 & 0.82$\pm$0.02 & --2.9$\pm$0.3 & 0.98$\pm$0.10 & --0.048 & 530$\pm$60 & --0.39$\pm$0.06 & --0.019$\pm$0.003\\ 
        12.02 & 0.7$\pm$0.02 & --5.9$\pm$0.6 & 0.93$\pm$0.10 & --0.057 & 590$\pm$60 & --0.70$\pm$0.10 & --0.040$\pm$0.006 \\
        14.35 & 0.56$\pm$0.02 & --5.9$\pm$0.6 & 0.75$\pm$0.08 & --0.069 & 580$\pm$60 & --0.72$\pm$0.11 & --0.050$\pm$0.007 \\
        16 & 0.45$\pm$0.02 & --4.0$\pm$0.4 & 0.66$\pm$0.07 & --0.079 & 580$\pm$60 & --0.49$\pm$0.07 & --0.039$\pm$0.006\\
        18.39 & 0.23$\pm$0.02 & & & -0.095 & & & \\ \hline
    \end{tabular}
    \caption{\label{tab:A-to-IC-2_Field}Relevant quantities and jumps, calculated changes of entropy, and calculated pressure dependencies for the discontinuous transition in magnetic field from A(SkL) to IC-2 for the $c$-axis according to Eq.~\eqref{eq:Clapeyron}. The column of ${\partial}B_{\mathrm{c}}/{\partial}T$ was calculated by taking the derivative of a polynomial fit to the values $B_{\mathrm{c}}(T)$.}
\end{table*}

\begin{table*}[ht]
    \centering
    \begin{tabular}{ccccccc}
    \hline \hline
        \multicolumn{7}{c}{A(SkL) $\rightarrow$ IC-2 -- $c$-axis: Relevant quantities, entropy changes and uniaxial pressure dependencies} \\
        $B$ (T) & $T_{\mathrm{c}}$ (K) & $\Delta$(dL/L) (10$^{-6}$) & $\Delta m$ ($\mu_{\mathrm{B}}$/Gd) & ${\partial}T_{\mathrm{c}}/{\partial}B$ (K/T) & $\Delta$S$_{\mathrm{calc}}$ (mJ/mol K) & $dT_{\mathrm{c}}/dp_{\mathrm{i}}$ (K/GPa) \\ \hline
        0 & 19.7$\pm$0.5 &  &  no SkL & & & \\
        0.1 & 19.3$\pm$0.3 &  & no SkL & --4.8 &   &  \\
        0.2 & 18.8$\pm$0.5 &  & no SkL & --8.1 &   &  \\
        0.3 & 17.9$\pm$0.7 &  & no SkL & --10.8 &  & \\
        0.4 & 16.4$\pm$0.4 & --5.0$\pm$1.0 & 0.36$\pm$0.04 & --13.0 & 310$\pm$40 & --1.2$\pm$0.3 \\
        0.5 & 15.3$\pm$0.4 & --7.4$\pm$1.0 & 0.54$\pm$0.04 & --14.8 & 410$\pm$40 & --1.3$\pm$0.2 \\
        0.6 & 13.6$\pm$0.3 & --7.0$\pm$0.7 & 0.70$\pm$0.04 & --16.0 & 490$\pm$30 & --1.02$\pm$0.12 \\
        0.7 & 12.0$\pm$0.3 & --5.1$\pm$0.5 & 0.77$\pm$0.04 & --16.7 & 510$\pm$30 & --0.70$\pm$0.08 \\
        0.8 & 10.4$\pm$0.3 & --2.7$\pm$0.3 & 0.86$\pm$0.04 & --17.0 & 560$\pm$30 & --0.34$\pm$0.04 \\
        0.9 & 8.60$\pm$0.3 & --1.3$\pm$0.1 & 0.91$\pm$0.04 & --16.7 & 610$\pm$30 & --0.15$\pm$0.02 \\ \hline 
    \end{tabular}
    \caption{\label{tab:A-to-IC-2_Temp}Relevant quantities and jumps, calculated changes of entropy, and calculated pressure dependencies for the discontinuous transition in temperature from A(SkL) to IC-2 for the $c$-axis according to Eq.~\eqref{eq:Clapeyron}. The column of ${\partial}T/{\partial}B$ was calculated by taking the derivative of a polynomial fit to the values T$_{\mathrm{c}}$(B), including the values from 0~T to 0.3~T.} 
\end{table*}

\begin{table*}[ht]
    \centering
    \begin{tabular}{cccccccc}
    \hline \hline
       \multicolumn{8}{c}{IC-2 $\rightarrow$ DP -- $c$-axis: Relevant quantities, entropy changes and uniaxial pressure dependencies} \\
        T (K) & B$_{\mathrm{c}}$ (T) & $\Delta(dL/L)$ (10$^{-6}$) & $\Delta m$ ($\mu_{\mathrm{B}}$/Gd) & ${\partial}B_{\mathrm{c}}/{\partial}T$ (T/K) & $\Delta S_{\mathrm{calc}}$ (mJ/mol K) & $dT_{\mathrm{c}}/dp_{\mathrm{i}}$ (K/GPa) & ${\partial}B_{\mathrm{c}}/{\partial}p_i$ (T/GPa) \\ \hline
        1.77 & 3.51$\pm$0.05 & --17.1$\pm$1.7 & 0.08$\pm$0.03 & --0.04$\pm$0.04 & 16$\pm$16 & --80$\pm$80 & --1.4$\pm$0.6 \\
        4.3 & 3.44$\pm$0.05 & --12.2$\pm$2 & 0.08$\pm$0.04 & --0.014$\pm$0.014 & 7$\pm$7 & --130$\pm$130 & --0.9$\pm$0.4 \\ 
        5.01 & 3.44$\pm$0.10 & --11.8$\pm$0.4 & 0.07$\pm$0.03 & --0.008$\pm$0.008 & 3$\pm$3 & --250$\pm$250 & --1.0$\pm$0.6 \\ 
        7.94 & 3.45$\pm$0.15 & --6$\pm$3 & 0.044$\pm$0.018 & 0.018$\pm$0.009 & -4$\pm$3 & 100$\pm$80 & -0.9$\pm$0.6\\ \hline
    \end{tabular}
    \caption{\label{tab:IC-2-to-DP} Relevant quantities and jumps, calculated changes of entropy, and calculated pressure dependencies for the discontinuous transition in magnetic field from IC-2 to DP for the $c$-axis according to Eq.~\eqref{eq:Clapeyron}. The column of ${\partial}B_{\mathrm{c}}/{\partial}T$ was calculated by taking the derivative of a polynomial fit to the values $B_{\mathrm{c}}(T)$.} 
\end{table*}

\begin{table*}
    \centering
    \begin{tabular}{ccccc}
    \hline \hline
       \multicolumn{5}{c}{IC-2 $\rightarrow$ fiFM -- $c$-axis: Relevant quantities, entropy changes and uniaxial pressure dependencies} \\
        T (K) & B$_{\mathrm{c}}$ (T) & ${\Delta}\lambda$ (10$^{-6}$/T) & ${\Delta}({\partial}M/{\partial}B)$ ($\mu_{\mathrm{B}}$/(T Gd)) & ${\partial}B_{\mathrm{c}}/{\partial}p_{\mathrm{i}}$ (T/GPa) \\ \hline
        9.99 & 3.7$\pm$0.1 & 15$\pm$2 & --0.15$\pm$0.03 & -0.7$\pm$0.2 \\ 
        12.02 & 3.0$\pm$0.1 & 17$\pm$2 & --0.19$\pm$0.03 & -0.6$\pm$0.2 \\
        14.35 & 2.3$\pm$0.1 & 19$\pm$2 & --0.25$\pm$0.03 & -0.5$\pm$0.1 \\ \hline
    \end{tabular}
    \caption{\label{tab:IC-2-to-fiFM} Relevant quantities and jumps and calculated field and pressure dependencies for the continuous transition from IC-2 to the field-induced ferromagnetic (fiFM) phase for the $c$-axis according to Eq.~\eqref{eq:Ehrenfest_1}, \eqref{eq:Ehrenfest_2}.} 
\end{table*}

\begin{table*}
    \centering
    \begin{tabular}{ccccc}
    \hline \hline
       \multicolumn{5}{c}{DP $\rightarrow$ fiFM -- $c$-axis: Relevant quantities, entropy changes and uniaxial pressure dependencies} \\
        T (K) & B$_{\mathrm{c}}$ (T) & ${\Delta}\lambda$ (10$^{-5}$/T) & ${\Delta}({\partial}M/{\partial}B)$ ($\mu_{\mathrm{B}}$/(T Gd)) & ${\partial}B_{\mathrm{c}}/{\partial}p_{\mathrm{i}}$ (T/GPa) \\ \hline
        1.77 & 8.8$\pm$0.5 & --4.7$\pm$1.0 & --0.159$\pm$0.018 & 1.9$\pm$0.5 \\ 
        4.3 & 7.94$\pm$0.4 & --1.9$\pm$1.0 & --0.094$\pm$0.009 & 1.3$\pm$0.7 \\ 
        5.01 & 6.6$\pm$0.4 & --1.4$\pm$1.0 & --0.074$\pm$0.018 & 1.2$\pm$0.9 \\ 
        7.94 & 6.48$\pm$0.6 & --0.7$\pm$0.4 & --0.065$\pm$0.018 & 0.6$\pm$0.4 \\ \hline
    \end{tabular}
    \caption{\label{tab:DP-to-fiFM} Relevant quantities and jumps and calculated field and pressure dependencies for the continuous transition from the depinned phase (DP) to the field-induced ferromagnetic (fiFM) phase for the $c$-axis according to Eq.~\eqref{eq:Ehrenfest_1}, \eqref{eq:Ehrenfest_2}.} 
\end{table*}

\begin{table*}[ht]
    \centering
    \begin{tabular}{ccccccc}
    \hline \hline
        \multicolumn{7}{c}{IC-1 $\rightarrow$ IC-1' -- $c$-axis} \\
        B & T$_{\mathrm{c}}$ & $\Delta$\ac & $\Delta$\cp & $\Delta ({\partial}M/{\partial}T)$ & $dT_{\mathrm{c}}/dp_i$ & $dT_{\mathrm{c}}/dB_i$ \\
        (T) & (K) & (10$^{-6}$) & (J/mol K) & (10$^{-3}\mu_{\mathrm{B}}$/f.u. K) & (K/GPa) & (K/T) \\ \hline
        0 & 12.8$\pm$0.7 & 4.1$\pm$0.5 & --2.7$\pm$0.5 &  & --1.4$\pm$0.3 & \\
        0.2 & 12.15$\pm$0.5 & 2.7$\pm$0.4 & --1.8$\pm$0.4 & --5.6$\pm$1.4 & --1.3$\pm$0.4 & --0.037$\pm$0.013 \\ 
        0.25 & 11.45$\pm$0.5 &  &  & --11$\pm$4 & & \\
        0.3 & 10.6$\pm$0.5 & 0.9$\pm$0.3 & --0.6$\pm$0.3 & --16$\pm$6 & --1.1$\pm$0.7 & --0.28$\pm$0.18 \\ 
        0.35 & 8.4$\pm$0.5 &  &  & --20$\pm$7 & & \\ \hline
    \end{tabular}
    \caption{\label{tab:T-star} Relevant quantities, jumps, and calculated field and pressure dependencies for the continuous transition at T$^*$ from IC-1 to IC-1' for the $c$-axis according to Eq.~\eqref{eq:Ehrenfest_1}, \eqref{eq:Ehrenfest_2}. Note that the \cp\ data was only measured at 0~T. The jumps in field $B$ were scaled by the changes of the jumps in \ac, i.e., $\Delta$\cp(B) = \cp(0) $\cdot~\Delta\alpha$(B)/$\Delta\alpha$(0).} 
\end{table*}

\end{document}